\documentclass[preprint]{elsarticle}
\usepackage{graphicx}
\usepackage[utf8]{inputenc}
\usepackage{amsmath}
\usepackage{verbatim}
\usepackage{amssymb}
\usepackage{listings}
\usepackage{color}
\usepackage{hyperref}
\usepackage{amsmath}
\usepackage{mathtools}
\usepackage{color}
\usepackage{xspace}

\usepackage{lineno}

\makeatletter
\DeclareFixedFont{\ttb}{T1}{txtt}{bx}{n}{\f@size} 
\DeclareFixedFont{\ttm}{T1}{txtt}{m}{n}{\f@size}  
\makeatother

\definecolor{deepblue}{rgb}{0,0,0.5}
\definecolor{deepred}{rgb}{0.6,0,0}
\definecolor{deepgreen}{rgb}{0,0.5,0}

\makeatletter
\newcommand\pythonstyle{\lstset{
language=Python,
basicstyle={\DeclareFixedFont{\temporary}{T1}{txtt}{m}{n}{\f@size}\temporary},
otherkeywords={self},             
keywordstyle=\ttb\color{deepblue},
emph={MyClass},          
emphstyle=\ttb\color{deepred},    
stringstyle=\color{deepgreen},
frame=tb,                         
showstringspaces=false            %
}}
\makeatother

\lstnewenvironment{python}[1][]
{
\pythonstyle
\lstset{#1}
}
{}

\newcommand\pythoninline[1]{{\pythonstyle\lstinline!#1!}}

\newcommand\subfig[2]{{Fig.~\ref{#1}{#2}}}
\newcommand\subcap[1]{{(#1):}}

\newcommand{\pFact}{p^{\text{Fact} }}
\newcommand{\ZZ}{\mathbb{Z}}

\newcommand{\partpart}[2]{\frac{\partial #1}{\partial #2}}

\newcommand{\Conf}{\ensuremath{c}}

\newcommand{\NewConf}{\ensuremath{c'}}

\newcommand{\LEG}[2]{(#1 \to #2)}
\newcommand{\FACTOR}[2]{\glb #1, #2 \grb}
\newcommand{\TYPE}[2][]{\gla \text{#2}\gra_{#1}}
\newcommand{\SETOFM}{\MCAL}

\newcommand{\BENDINGTYPE}{\text{bending}} 
\newcommand{\LJTYPE}{\text{LJ}} 
\newcommand{\COULOMBTYPE}{\text{Coulomb}} 
\newcommand{\NACT}{n_{\text{ac}}} 
\newcommand{\NPROCESS}{n_{\text{pr}}} 

\newcommand{\JF}{\textsc{JF}\xspace}
\newcommand{\JFL}{\textsc{JeLLyFysh}\xspace}

\newcommand{\JFA}{\JF application\xspace}
\newcommand{\JFV}{\textsc{JeLLyFysh}-Version1.0\xspace}
\newcommand{\JFVS}{\textsc{JF-V1.0}\xspace}
\newcommand{\JFVTWO}{\JF (Version 2.0)\xspace}
\newcommand{\pp}{point mass\xspace}

\newcommand{\pps}{point masses\xspace}
\newcommand{\PPS}{Point masses\xspace}
\newcommand{\cpp}{composite point object\xspace}
\newcommand{\cpps}{composite point objects\xspace}

\newcommand{\onecircle} {\raisebox{.5pt}{\textcircled{\raisebox{-.9pt} {1}}}}
\newcommand{\twocircle} {\raisebox{.5pt}{\textcircled{\raisebox{-.9pt} {2}}}}
\newcommand{\threecircle} {\raisebox{.5pt}{\textcircled{\raisebox{-.9pt} {3}}}}
\newcommand{\fourcircle} {\raisebox{.5pt}{\textcircled{\raisebox{-.9pt} {4}}}}
\newcommand{\fivecircle} {\raisebox{.5pt}{\textcircled{\raisebox{-.9pt} {5}}}}
\usepackage{bold-extra}

\newcommand{\COMMANDLINE}[1]{\texttt{\lstinline!#1!}}
\newcommand{\CLASS}[1]{\pythoninline{ #1}}
\newcommand{\OBJECT}[1]{\pythoninline{ #1}}
\newcommand{\STAGE}[1]{\pythoninline{ #1}}
\newcommand{\VARIABLE}[1]{\pythoninline{ #1}}
\newcommand{\INI}[1]{\pythoninline{#1}}
\newcommand{\TAG}[1]{\pythoninline{#1}}
\newcommand{\METHOD}[1]{\pythoninline{#1}}
\newcommand{\MODULE}[1]{\pythoninline{#1}}
\newcommand{\PACKAGE}[1]{\pythoninline{#1}}
\newcommand{\PATH}[1]{\pythoninline{#1}}
\newcommand{\PROPERTY}[1]{\pythoninline{#1}}

\newcommand{\SCRIPT}[1]{\texttt{#1}}
\newcommand{\SECTION}[1]{\pythoninline{[#1]}}

\newcommand{\LOGLEVEL}[1]{\pythoninline{#1}}
\newcommand{\via}{\emph{via}\xspace}
\newcommand{\ident}{\sigma}

\newcommand{\SET}[1]{\{#1\}}

\newcommand{\eq}[1]{eq.~\eqref{#1}}

\newcommand{\fig}[1]{Fig.~\ref{#1}}
\newcommand{\quot}[1]{``#1''}

\newcommand{\sect}[1]{Section~\ref{#1}} 
\newcommand{\secttwo}[2]{Sections~\ref{#1} and~\ref{#2}}

\newcommand{\MCAL}{\mathcal{M}}  
\newcommand{\OCAL}{\mathcal{O}}  
\newcommand{\expa}[1]{\mathrm{e}^{#1}}   
\newcommand{\expb}[1]{\exp \glb #1 \grb} 
\newcommand{\expc}[1]{\exp \glc #1 \grc} 
\newcommand{\cosb}[2][]{\cos^{#1} \glb #2 \grb}  

\newcommand{\gla}{\,}  
\newcommand{\gra}{}  
\newcommand{\glb}{\left(}  
\newcommand{\grb}{\right)}  
\newcommand{\glc}{\left[}  
\newcommand{\grc}{\right]}  

\newcommand{\const}{\text{const}}

\newcommand{\TO}{,\ldots,}
\newcommand{\VEC}[1]{\mathbf{#1}}
\newcommand{\Lvec}{\VEC{L}}
\newcommand{\mvec}{\VEC{m}}
\newcommand{\nvec}{\VEC{n}}

\newcommand{\qvec}{\VEC{q}}
\newcommand{\rvec}{\VEC{r}}

\newcommand{\svec}{\VEC{s}}

\newcommand{\vvec}{\VEC{v}}

\newcommand\bigO[1]{\ensuremath{\OCAL(#1)}}

\begin{document}
\date{\today}
\title{\JFV\ - a Python application for all-atom event-chain Monte Carlo}
\author[LPENS,BONN]{Philipp~Höllmer}
\author[LPENS]{Liang~Qin}
\author[BRISTOL]{Michael~F.~Faulkner}
\author[ESPCI]{A.~C.~Maggs}
\author[LPENS,MPIPKS]{Werner~Krauth\corref{vvv}}
\cortext[vvv]{Corresponding author, email address:
\pythoninline{werner.krauth@ens.fr}}

\address[LPENS]{Laboratoire de Physique de l’Ecole normale supérieure, ENS, 
Université PSL, CNRS, Sorbonne Université, Université Paris-Diderot, Sorbonne 
Paris Cité, Paris, France}

\address[BONN]{Bethe Center for Theoretical Physics, University of Bonn, 
Nussallee 12, 53115 Bonn, Germany}

\address[BRISTOL]{H.~H. Wills Physics Laboratory, University of Bristol, 
Tyndall Avenue, Bristol BS8 1TL, United Kingdom}
\address[ESPCI]{CNRS UMR7083, ESPCI Paris, PSL Research University, 10 rue
Vauquelin, 75005 Paris, France}

\address[MPIPKS]{Max-Planck-Institut für Physik komplexer Systeme, Nöthnitzer 
Str. 38, 01187 Dresden, Germany}

\begin{abstract}
We present \JFV, an open-source Python application for event-chain Monte Carlo 
(ECMC), an event-driven irreversible Markov-chain Monte Carlo algorithm for
classical $N$-body simulations in statistical mechanics, biophysics and 
electrochemistry. The application's architecture closely mirrors the 
mathematical formulation of ECMC. Local potentials, long-ranged Coulomb 
interactions and multi-body bending potentials are covered, as well as 
bounding potentials and cell systems including the cell-veto algorithm. 
Configuration files illustrate a number of specific implementations for 
interacting atoms, dipoles, and water molecules.
\end{abstract}

\maketitle

\section{Introduction}
\label{sec:Introduction}
\begin{figure}[htb]
\begin{center}
\includegraphics{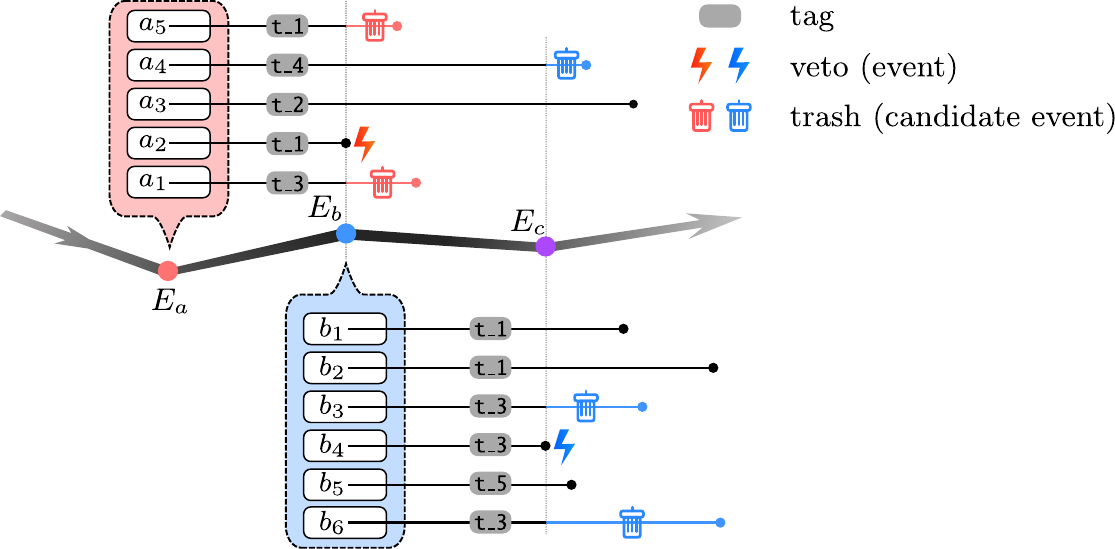}
\end{center}
\caption{ECMC time evolution. At 
events $E_a, E_b, E_c, \dots$, a number of factors 
($\SET{a_1, a_2 \TO a_5 }, \SET{b_1, b_2 \TO b_6 }, \ldots$) are 
activated. For each leg ($\LEG{E_a}{E_b}$, $\LEG{E_b}{E_c}$, \dots), each 
factor 
must at all times independently accept the continued non-interacting evolution, 
and must determine a candidate event time at which this is no longer the case. 
The earliest candidate event time (which determines the veto) and its 
out-state yield the next event (the event $E_b$ is  triggered by
$a_2$). 
In \protect \JFVS, 
after committing an event to the global state, 
candidate events with certain tags are trashed 
(tags \protect \TAG{t_1}, \protect \TAG{t_3} at $E_b$) or maintained active 
(tags
\protect \TAG{t_2}, \protect \TAG{t_4} at $E_b$), and others are newly 
activated.
\JF introduces 
non-confirmed events and also pseudo-factors, which complement the factors 
of ECMC, and which may  also trigger events. }
\label{fig:EventFlow}
\end{figure}
Event-chain Monte Carlo (ECMC) is an irreversible continuous-time Markov-chain 
algorithm~\cite{Bernard2009,Michel2014JCP} that often equilibrates faster than 
its reversible 
counterparts~\cite{Nishikawa2015,KapferKrauth2017,Lei2018,Lei2018b,Lei2019}. 
ECMC has been successfully applied to the classic $N$-body all-atom problem in 
statistical physics~\cite{Bernard2011,Kapfer2015PRL}. The algorithm implements 
the time evolution of a piecewise non-interacting, deterministic, 
system~\cite{BierkensPDMC2017}. Each straight-line, non-interacting leg of this 
time evolution terminates in an event, defined through the  event time at which 
it takes place and through the out-state, the updated starting configuration for 
the ensuing leg. An event is chosen as the earliest of a set of candidate 
events, each of which is sampled using information contained in a so-called 
factor. The entire trajectory samples the equilibrium probability distribution.

ECMC departs from virtually all Monte Carlo methods in that it does not
evaluate the equilibrium probability density (or its ratios). In statistical 
physics, ECMC thus
computes neither the total potential (or its changes) nor the total force on 
individual
\pps.  Rather, the decision to continue on the current leg of the 
non-interacting time
evolution builds on a consensus which is established through the factorized
Metropolis algorithm~\cite{Michel2014JCP}. A veto puts an end to the consensus,
triggers the event, and terminates the leg
(see \fig{fig:EventFlow}). In the
continuous problems for which ECMC has been conceived, the veto is caused by a
single factor.

The resulting event-driven ECMC algorithm is reminiscent of 
molecular dynamics, and in particular of event-driven molecular 
dynamics~\cite{Alder1957,AlderWainwright1959,BannermanLue2010}, in that there 
are 
velocity vectors (which appear as lifting variables). 
These velocities do not correspond to the physical (Newtonian) dynamics of the 
system. ECMC differs  from 
molecular dynamics in three respects: First, 
ECMC is event-driven, and it remains approximation-free, for any 
interaction
potential~\cite{Peters_2012}, 
whereas event-driven molecular dynamics is restricted to hard-sphere or 
piecewise constant potentials. (Interaction potentials in biophysical 
simulation codes have been coarsely discretized~\cite{DingTsaoDokholyan2008} in 
order to fit into the event-driven 
framework~\cite{ShirvanyantsDingDokholyan2012,ProctorDingDokholyan2011,Proctor2016}.) 
Second, in ECMC, most \pps are at rest at any time, whereas in molecular 
dynamics, all \pps typically have non-zero velocities. In ECMC, an 
arbitrary fixed number of (independently) active \pps (with non-zero 
velocities) and identical velocity vectors for all of them  may be chosen. In 
\JFVS, as in most previous applications of ECMC, only a single independent 
\pp is active. The ECMC  dynamics is thus very  simple, yet it mixes and 
relaxes at a  rate at least as fast as in molecular 
dynamics~\cite{KapferKrauth2017,Lei2018b,Lei2019}.  Third, ECMC by 
construction exactly 
samples the Boltzmann (canonical) distribution, whereas molecular dynamics is 
in 
principle micro-canonical, that is, energy-conserving. Molecular dynamics is 
thus  generally coupled to a thermostat in order to yield the Boltzmann 
distribution. The thermostat there also  eliminates 
drift in physical observables due to integration errors. 
ECMC is free from truncation and discretization errors.

ECMC samples the equilibrium Boltzmann distribution without being itself in 
equilibrium, as it violates the detailed-balance condition.
Remarkably, it establishes the aforementioned consensus and proceeds 
from one event to the next with \bigO{1} computational effort even for 
long-range potentials, as was demonstrated for soft-sphere models, the Coulomb 
plasma~\cite{KapferKrauth2016,KapferKrauth2017}, and for the simple 
point-charge 
with flexible water molecules (SPC/Fw) 
model~\cite{WuTepperVoth2006,Faulkner2018}. 

\JFL (\JF) is a general-purpose Python application that implements ECMC for a 
wide range of physical systems, from \pps interacting with central potentials  
to \cpps such as finite-size dipoles, water molecules, and eventually peptides 
and polymers. The application's architecture closely mirrors the mathematical 
formulation that was presented previously (see~\cite[Sect II]{Faulkner2018}). 
The application can run on virtually any computer, but it also allows for 
multiprocessing and, in the future, for parallel implementations. It is being 
developed as an open-source project on GitHub. Source code may be forked, 
modified, and then merged back into the project (see \sect{sec:LicGitHubPython} 
for access information and licence issues). Contributions to the application are 
encouraged. 

The present paper introduces the general architecture and the key features of 
\JF. It accompanies the first public release of the application, \JFV (\JFVS). 
\JFVS implements ECMC for homogeneous, translation-invariant $N$-body systems in 
a regularly shaped periodic simulation box and with interactions that can be 
long-ranged. 
In addition, the present paper presents a cookbook 
that illustrates the  application for simplified core examples that can be run 
from configuration files and validated against published 
data~\cite{Faulkner2018}. A full-scale simulation benchmark against the Lammps 
application is published elsewhere~\cite{Qin2019}.

The \JF application presented in this paper is intended to grow into 
a basis code 
that will foster the development of irreversible Markov-chain algorithms and 
will apply to a wide range of 
computational problems, from statistical physics to field 
theory~\cite{HasenbuschSchaefer2018}. It may prove useful in domains that have 
traditionally been reserved to molecular dynamics, and in particular in the  
all-atom Coulomb problem in biophysics and electrochemistry.

The content of the present paper is as follows: The remainder of 
\sect{sec:Introduction} discusses the general setting of \JF as it implements 
ECMC. \sect{sec:Architecture} describes its mediator-based 
architecture~\cite{GammaDesignPatterns1994}. \sect{sec:EventHandlers}
discusses how the eponymous events of ECMC are determined in the event handlers 
of \JF. \sect{sec:System} presents system definitions and tools, 
such as the user interface realized through configuration files, 
the simulation box, the cell systems, and the interaction potentials. 
\sect{sec:Cookbook}, the cookbook, discusses a number of 
worked-out examples for previously presented systems of atoms, dipoles or water 
molecules with Coulomb interactions~\cite{Faulkner2018}. 
\sect{sec:LicGitHubPython} discusses licence issues, code availability and code 
specifications. \sect{sec:Conclusions} presents an outlook on essential 
challenges and a preview of future releases of the application.

\subsection{Configurations, factors, pseudo-factors, events, event handlers}
\label{sec:IntroductionFactorsPseudoFactors}

In ECMC, configurations $\Conf = \SET{\svec_1 \TO  \svec_i \TO \svec_N}$ are 
described by continuous time-dependent variables where $\svec_i(t)$
represents 
the position of the $i$th of $N$  \pps (although it may also stand for the 
continuous angle of a spin on a lattice~\cite{Nishikawa2015}). \JF is 
an event-driven implementation of ECMC, and it treats \pps and certain 
collective variables (such as the barycenter of a \cpp) on an equal footing. 
Rather than the time-dependent variables $\svec_i(t)$, its fundamental 
particles (\OBJECT{Particle} objects) are individually time-sliced positions 
(of 
the \pps or \cpps).  Non-zero velocities and time stamps are also recorded, 
when 
applicable. The full information can be packed into units (\OBJECT{Unit} 
objects), that are
moved around the application (see \sect{sec:IntroductionGlobalInternal}).

Each configuration $\Conf$ has a total potential 
$U(\SET{\svec_1 \TO \svec_N})$, and its equilibrium probability density 
$\pi$ is given by the Boltzmann weight
\begin{equation}
\pi(\SET{\svec_1 \TO \svec_N}) = \expc{-\beta U(\SET{\svec_1 \TO \svec_N})},
\label{equ:BoltzmannWeight}
\end{equation}
that is sampled by ECMC (see~\cite{Faulkner2018}). The total potential
$U$ is decomposed as
\begin{equation}
 U(\SET{\svec_1 \TO \svec_N}) = \sum_{M \in \SETOFM}
U_M (\SET{\svec_i: i \in I_M}),
\label{equ:PotentialFactorized}
\end{equation}
and the Boltzmann weight of \eq{equ:BoltzmannWeight} is written as a product 
over terms that depend on factors $M$,
with their corresponding factor potentials $U_M$.
A factor $M= (I_M, T_M)$ consists of an index set $I_M$ and of a factor 
type $T_M$, and $\SETOFM$ is the set of factors that have a non-zero 
contribution to \eq{equ:PotentialFactorized} for some configuration $\Conf$. In 
the SPC/Fw water model, for example, one factor $M$ with factor type $T_M = 
\COULOMBTYPE$ might describe all the Coulomb potentials between two given water 
molecules, and the factor index set $I_M$ would contain the identifiers 
(indices) of the 
involved four hydrogens and two oxygens (see \sect{sec:CookbookSPC}). 

\begin{figure}[htb]
\begin{center}
\includegraphics{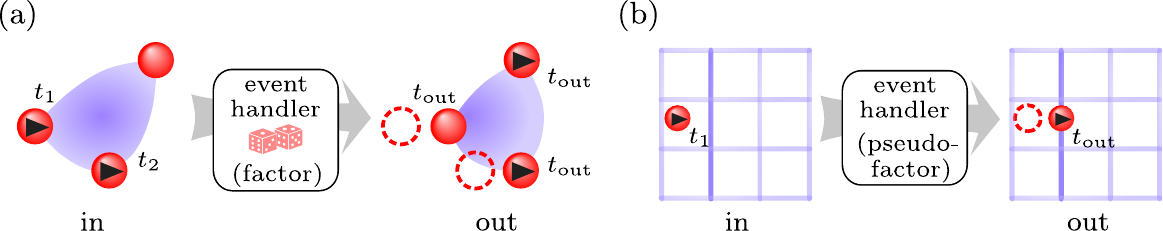} 
\end{center}
\caption{Factors and pseudo-factors. \subcap{a} In-state and sampled  out-state
(each with two active units) for a  three-unit factor $M$ (implementing, for 
example, the inter-molecular bending potential $U_M$ of 
\sect{sec:SystemPotentialsBending}). \subcap{b} In- and out-states for a 
cell-boundary event handler realizing a pseudo-factor. Times at which units are 
time-sliced are indicated. $t_\text{out}$ is the event time.}
\label{fig:FactorPseudoFactor}
\end{figure}

ECMC relies on the factorized Metropolis algorithm~\cite{Michel2014JCP}, where 
the  move from a configuration $\Conf$ to another one, $\NewConf$,  is accepted 
with probability
\begin{equation}
    \pFact(\Conf \to \NewConf) = \prod_{M} \min \glc 1,
\expb{-\beta \Delta  U_{M}} \grc,
\label{equ:MetropolisFactorized}
\end{equation}
where $\Delta     U_{M} = U_{M}(\NewConf_M)  - U_M( \Conf_M)$.
Rather than to evaluate the right-hand side of \eq{equ:MetropolisFactorized}, 
the product over the factors is interpreted as corresponding to a conjunction 
of independent Boolean random variables
\begin{equation}
    X^{\text{Fact}}(\Conf \to \NewConf)  = \bigwedge_{M \in \SETOFM}
X_{M}(\Conf_M \to \NewConf_{M}).
\label{equ:MetropolisFactorizedBoolean}
\end{equation}
In this equation, $X^{\text{Fact}}(\Conf \to \NewConf)$ is \quot{True} (the 
proposed Monte Carlo move is accepted) if the independently sampled factorwise 
Booleans $X_{M}$ are all \quot{True}. Equivalently, the move $\Conf \to 
\NewConf$ is accepted if it is independently accepted by all factors. 
This realizes the aforementioned consensus decision (see \fig{fig:EventFlow}). 
For an infinitesimal displacement, the random variable $X_M$ of only a single 
factor $M$ can be \quot{False}, and the factor $M$ vetoes the consensus, 
creates an event, and starts a new leg. 
In this process, $M$ requires only the 
knowledge of the factor in-state (based on the configuration 
$\Conf_M$, 
and the information on the move), 
and the factor out-state (based on $\Conf'_M$)
provides all information on the evolution of the system after the event.
The event is needed in order to enforce the global-balance condition (see 
\subfig{fig:FactorPseudoFactor}{a}). 
In this process, lifting variables~\cite{Diaconis2000}, corresponding 
to generalized velocities, allow one to repeat moves of the same type (same 
particle, same displacement), as long 
as 
they are accepted by consensus.\footnote{For
concreteness, the 
lifting variables  in this paper are referred to as \quot{velocities}, 
although they are not derived from mechanical equations of motions and their
conservation  laws. The concept of lifting variables is more 
general~\cite{Diaconis2000}.} Physical and lifting variables build the 
overcomplete description of the Boltzmann distribution at the base of 
ECMC, and they correspond to the global physical and 
global lifting states of \JF, its global state.

\JF, the computer application, is entirely formulated in terms of events, beyond 
the requirements of the implemented event-driven ECMC algorithm. The application 
relies on the concept of pseudo-factors, which complement the factors in 
\eq{equ:PotentialFactorized}, but are independent of potentials and without 
incidence on the global-balance condition (see 
\subfig{fig:FactorPseudoFactor}{b}). In \JF, the sampling of configuration 
space, for example, is expressed through events triggered by pseudo-factors. 
Pseudo-factors also trigger events that interrupt one continuous motion (one 
\quot{event chain}~\cite{Bernard2009}) and start a new one. Even the start and 
the 
end of each run of the application are formulated as events triggered by 
pseudo-factors.

In ECMC,  among all factors $M$ in \eq{equ:PotentialFactorized}, only those  for 
which $U_M$ changes along one leg can trigger events. In \JF, these factors are 
identified in a separate element of the application, the activator (see 
\sect{sec:ArchitectureActivator}), and they are realized in yet other elements, 
the event handlers. An event handlers may require an in-state. It then 
computes 
the candidate event time and its out-state (from the in-state, from the factor 
potential, and from random elements). The complex operation of the activator and 
the event handlers is organized in \JFVS with the help of a tag activator, with 
tags essentially providing finer distinction than the factor types $T_M$. A 
tagger identifies a certain pool of factors, and also singles out factors that 
are to be activated for each tag. The triggering of an event associated with a 
given tag 
entails the trashing of candidate events with certain tags, while other 
candidate events are maintained (see \fig{fig:EventFlow}). Also, new candidate 
events have to be computed by event handlers with given tags. 
This entire process is managed by the tag activator. 

\subsection{Global state, internal state}
\label{sec:IntroductionGlobalInternal}

In the 
event-driven formulation of ECMC, a \pp with identifier $\ident$
and with zero velocity  is simply represented through 
its position, while an active \pp (with non-zero velocity) is represented
through a time-sliced position $\svec_\ident(t_\ident)$, 
a time stamp $\ident(t_\ident)$ and a velocity $\vvec_\ident$: 
\begin{equation}
     \svec_\ident(t) = \begin{cases}
\svec_\ident & \quad \text{if $\vvec_\ident = 0$} \\
\svec_\ident(t_\ident) + (t - t_\ident) \vvec_\ident & \quad \text{else (active 
\pp)}
                       \end{cases}. 
\label{equ:EventDrivenPosition}     
\end{equation}
An active \pp thus requires storing of a velocity $\vvec_i $ and of a time 
stamp 
$t_\ident$, in addition to the time-sliced position $\svec_\ident(t_\ident)$. In 
\JF, the global state traces all the information in 
\eq{equ:EventDrivenPosition}. It is broken up into the global physical state, 
for the time-sliced positions $\svec_\ident$,  and the global lifting state, for 
the non-zero velocities $ \vvec_\ident$ and the time stamps $t_\ident$.

\begin{figure}[htb] \begin{center} 
\includegraphics{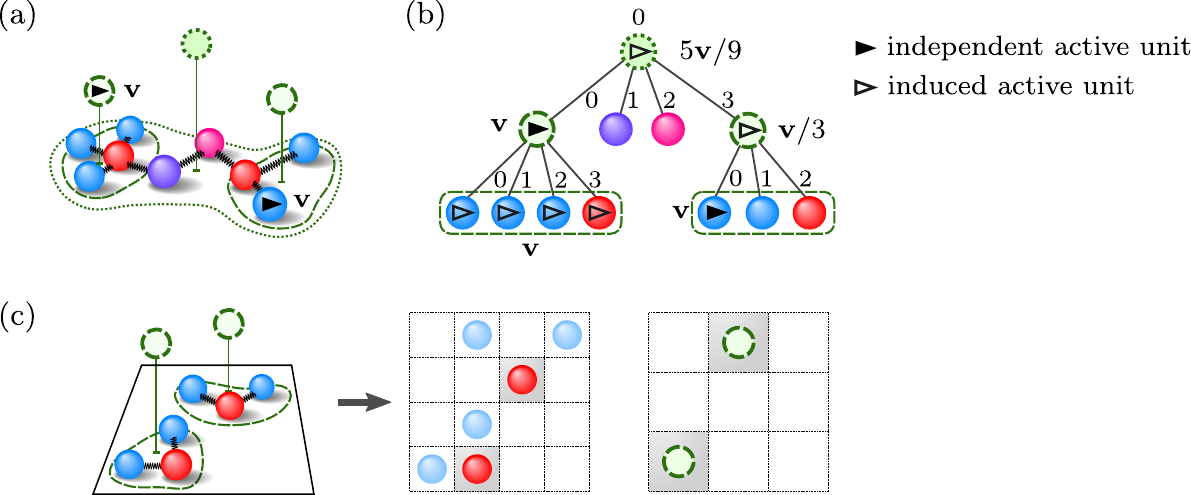}
\end{center}
\caption{Tree representation of \cpps in \JFVS. \subcap{a} Molecule with 
functional parts. \subcap{b} Tree representation, with leaf nodes 
 for the individual atoms and higher-level nodes for  barycenters.
 Nodes each have a particle (a \protect \OBJECT{Particle} 
 object) containing a position vector and charge values. 
  A unit (a \protect \OBJECT{Unit} object), associated with a node,
 copies out the particle's identifier and its complete global-state 
 information.
 \subcap{c} Internal representation of 
\cpps with separate cell systems for particle identifiers on different 
levels. On the leaf level, only one kind of particles is tracked. }
\label{fig:CompositeParticleCells}
\end{figure}

\JF represents \cpps as trees described by nodes. Leaf nodes correspond to 
the individual \pps. A tree's inner nodes may represent, for example, the 
barycenters of parts of a molecule, and the root node that of the entire 
molecule (see 
\subfig{fig:CompositeParticleCells}{a-b}). The velocities inside a \cpp are kept 
consistent, which means that the global lifting state includes non-zero 
velocities and time stamps of inner and root nodes. 
The storing element of the 
global state in \JF is the state handler (see 
\sect{sec:ArchitectureStateHandler}). The global state is not directly accessed 
by other elements of the application, but branches of the tree can be extracted 
(copied) temporarily, together with their unit information.
Independent and induced units differentiate between those that appear in ECMC 
and those that are carried along in order to assure consistency (see 
\fig{fig:CompositeParticleCells}).

For internal computations, the global state may be supplemented by an 
internal state that is kept, not in the state handler, but in the activator part 
of the application (see \sect{sec:ArchitectureActivator}). 
In \JFVS, the internal state consists in cell-occupancy systems, which 
associate identifiers of \cpps or \pps to cells. (An identifier is a 
generalized particle index with, in the case of a tree, a number of elements 
that correspond to the level of the corresponding node.)
In \JF, 
cell-occupancy systems are used for book-keeping, and also 
for cell-based bounding potentials.
\JFVS 
requires consistency between the time-sliced particle information and the units.
This means that the time-sliced position $\svec_\ident(t_\ident)$ and the 
time-dependent position $\svec_\ident(t)$ in 
\eq{equ:EventDrivenPosition} 
belong to the same cell (see \subfig{fig:FactorPseudoFactor}{b}). 
Several 
cell-occupancy systems may coexist within the internal state (possibly on 
different tree-levels and with different cell systems, see 
\subfig{fig:CompositeParticleCells}{c} and 
\sect{sec:CookbookSPCCVCV}).
ECMC requires time-slicing only for units 
whose velocities are modified. Beyond the consistency requirements, 
\JFVS performs time-slicing  also for unconfirmed events, 
that is, for 
triggered events for which, after all, the out-state continues the 
straight-line motion of the in-state (see 
\sect{sec:EventHandlerFactorsBound}). 

\subsection{Lifting schemes}
\label{sec:IntroductionLifting}
In its lifted representation of the Boltzmann distribution, ECMC introduces 
velocities for which there are many choices, that is, lifting schemes. The 
number of independent active units can in particular be set to any value $\NACT 
> 1$ and then held fixed throughout a given run. This generalizes easily from 
the known $\NACT=1$ case~\cite{Harland2017}. A simple $\NACT$-conserving 
lifting 
scheme uses a factor-derivative table (see~\cite[Fig. 2]{Faulkner2018}), but 
confirms the active out-state unit only if the corresponding unit is not active 
in the in-state (its velocity is \pythoninline{None}). For $|I_M| > 3$, the 
lifting scheme (the way of determining the out-state given the in-state) is not 
unique, and its choice influences the ECMC dynamics~\cite{Faulkner2018}. In 
\JFVS, different lifting-scheme classes are provided in the \JF\ 
\PACKAGE{lifting} package. They all construct independent-unit out-state 
velocities for independent units that equal the in-state velocities. This 
appears as the most natural choice in spatially homogeneous 
systems~\cite{Bernard2009}.

\subsection{Multiprocessing}
\label{sec:IntroductionMultiprocessing}

In ECMC, factors are statistically independent. 
In \JF, therefore, the event handlers that realize these factors can be run 
independently on a multiprocessor machine. 
With multiprocessor support enabled, candidate events  
are concurrently determined by event handlers on separate processes, using the 
Python \MODULE{multiprocessing} module.  Candidate event times are then first 
requested in parallel from active event handlers, and then the out-state for 
the 
selected event. Given a sufficient number of available processors, 
out-states 
may be computed for candidate events in advance, before they are requested 
(see 
\sect{sec:ArchitectureMediator}). The event handlers themselves correspond to 
processes that usually last for the entire duration of one ECMC run. When not 
computing, event handlers are either in \STAGE{idle} stage waiting to compute 
an candidate event 
time or in \STAGE{suspended} stage waiting to compute an out-state. 

Using multiple processes instead of threads circumvents the Python global 
interpreter lock, but the incompressible time lag due to data exchange slows 
down the multiprocessor implementation of the mediator with respect to the 
single-processor implementation. 

\subsection{Parallelization}
\label{sec:IntroductionParallelization}

\begin{figure}[htb]
\begin{center}
\includegraphics{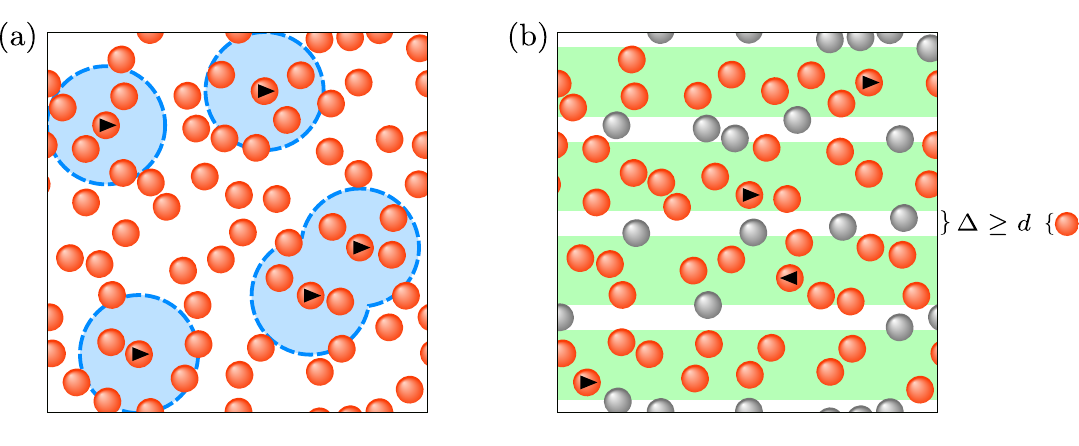}
\end{center}
\caption{Parallel ECMC with local potentials (interaction range $d$). 
\subcap{a} Multiprocess version with $\NACT \ll N$ active units. Nearby 
active units avoid conflict in a single process.
\subcap{b} Domain decomposition with separated stripes. Particles in between
stripes are immobile. The separation region (of width $\Delta$) is wider than 
$d$, so that all conflict between stripes is avoided 
(see~\cite{KapferPolytope2013}).}
\label{fig:ParallelSchemes}
\end{figure}

ECMC generalizes to more than one independent active unit, and a sequential, 
single-process ECMC computation remains trivially correct for arbitrary  $\NACT$ 
(although \JFVS only fully implements the $\NACT=1$ case). The relative 
independence of a small number of independent active units in a large system, 
for $1 \ll \NACT \ll N$, allows one to consider the simultaneous committing in 
different processes of $\NPROCESS$ events to the shared global state. (A 
conflict 
arises if this disagrees with what would result by committing them in a single 
process.) 
If $\NPROCESS \ll N$, conflicts between processes disappear (for short-range 
interacting systems) if nearby active units are treated in a single process (see 
\subfig{fig:ParallelSchemes}{a}). The parallel implementation of ECMC, for 
short-range interactions, is conceptually much simpler than that of 
event-driven 
molecular dynamics~\cite{MillerLuding2003,Herbordt2009,KhanHerbordt2011}, and it 
may well extend to long-range interacting system.

An alternative type of  parallel ECMC, domain decomposition into $\NACT$ 
stripes, was demonstrated for two-dimensional hard-spheres systems, and 
considerable speed-up was reached~\cite{KapferPolytope2013}. Here, stripes are 
oriented parallel to the 
velocities, with one active unit per stripe. Stripes are 
isolated 
from each other by immobile layers of spheres~\cite{KapferPolytope2013}, which 
however cause rejections (or reversals of one or more components of the 
velocity). The stripe decomposition eliminates all scheduling conflicts.
As any domain 
decomposition~\cite{MillerLuding2003}, it is restricted to physical models with 
short-range interactions. It is not implemented in \JFVS (see 
\subfig{fig:ParallelSchemes}{b}).

\section{\JF architecture}
\label{sec:Architecture}
\JF adopts the design pattern based on a 
mediator~\cite{GammaDesignPatterns1994}, which serves as the central hub for
the other elements that do not directly connect to each other. In this 
way, interfaces and data exchange are particularly simple. The mediator design 
maximizes modularity in view of future extensions of the application.

\subsection{Mediator}
\label{sec:ArchitectureMediator}

The mediator is doubled up into two modules (with \CLASS{SingleProcessMediator} 
and \CLASS{MultiProcessMediator} classes). The \METHOD{run} method of either 
class is called by the executable \pythoninline{run.py} script of the 
application, and it loops over the legs of the continuous-time evolution. The 
loop is interrupted when an \pythoninline{EndOfRun} exception is raised, and a 
\METHOD{post_run} method is invoked. For the single-process mediator, all the 
other elements are instances of 
classes that provide public methods. In particular, the mediator interacts with 
event handlers. For the multi-process 
mediator, each event handler has its own autonomous iteration loop and runs 
in a separate process. It exchanges data with the mediator through a two-way 
pipe. Receiving ends on both sides detect when data is available using 
the pipe's \METHOD{recv} methods.

In \JFVS, the same event-handler classes are used for the single-process and 
multi-process mediator classes. The multi-process mediator  achieves this
through a monkey-patching technique. It dynamically adds a
\METHOD{run_in_process} method to each created instance of 
an event handler, which then runs as an autonomous iteration loop in a process 
and reacts to shared flags set by the mediator. The multi-process mediator 
in addition decorates the event handler's \METHOD{send_event_time} and 
\METHOD{send_out_state} methods so that output is not 
simply returned (as it is in the single-process mediator) but rather 
transmitted through a pipe. Only the mediator accesses the event handlers, and 
these re-definitions of methods and classes (which abolish the need for
two versions for each event-handler class) are certain not to produce undesired 
side 
effects. 

\begin{figure}[htb]
\begin{center}
\includegraphics{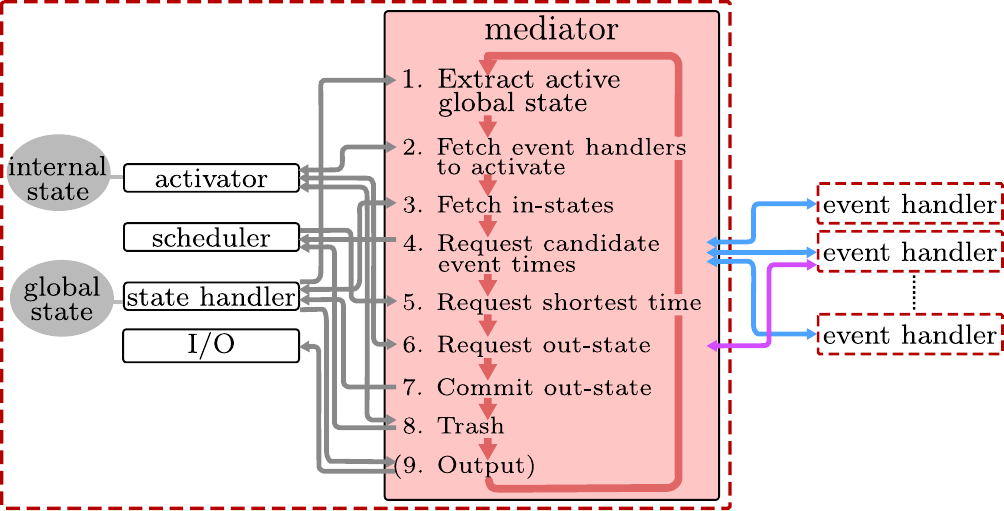}
\end{center}
\caption{\JF architecture, built on the mediator design pattern. The iteration 
loop takes the system from one event to the next (for example from $E_a$ to 
$E_b$ in \fig{fig:EventFlow}). All elements of \JF interact with the 
mediator, but not with each other. The multi-process mediator 
interacts with event handlers running on separate processes, and exchanges data 
\via pipes.}
\label{fig:MediatorSingle}
\end{figure}

On one leg of the continuous-time evolution, the mediator goes through nine 
steps (see \fig{fig:MediatorSingle}). In  step $1$, the active global state 
(that part of the global state that appears in the global lifting state) is 
obtained from the state handler. (In the tree state handler of \JFVS, branches 
of independent units are created for all identifiers that appear in the 
lifting state.) Knowing the preceding event handler (which initially is 
\pythoninline{None}) and  the active global state, it then obtains from the 
activator, in step $2$, the event handlers to activate together with their 
in-state identifiers. For this, the activator may rely on its internal state, 
but not on the global state, to which it has no access. In step $3$,  the 
corresponding in-states are extracted (that is, copied) from the state handler. 
In step $4$, candidate event times are requested  from the appropriate event 
handlers and pushed into the scheduler's \METHOD{push_event} method. In step 
$5$, the mediator obtains the earliest candidate event time from the 
scheduler's \METHOD{get_succeeding_event} method and and asks its event handler 
for 
the event out-state (step $6$) to be committed to the global state (step $7$). 
The activator,  in step $8$, determines which  candidate events are to be 
trashed (in \JFVS: based on their tags), that is, which candidate event times 
are to be eliminated from the scheduler. Also, the activator collects the 
corresponding event handlers, as they become available to determine new 
candidate events. In the optional final step $9$, the mediator may connect
(\via the input--output handler) to an output handler, depending on the 
preceding event handler. A mediating method defines the arguments sent to the 
output handler (for example the extracted global state), and considerable 
computations may take place there. 

The multi-process mediator uses a single pipe to receive the candidate 
event time and the out-state from an 
event handler. In order to distinguish the received 
object, the mediator assigns four different stages to the event handlers 
(\STAGE{idle}, \STAGE{event_time_started}, \STAGE{suspended}, 
\STAGE{out_state_started} stages). The assigned stage determines which flags 
can be set to 
start the \METHOD{send_event_time} or \METHOD{send_out_state} methods. 
It also determines the nature of the data contained in the pipe. 
In the \STAGE{idle} stage, the mediator can set the  starting flag  after which 
the event handler will wait to receive the in-state through the pipe. This 
starts the \STAGE{event_time_started} stage during which the event handler
determines the next candidate event time and places it into the pipe. After the 
mediator has recovered the data from the pipe, it places the event handler into 
the \STAGE{suspended} stage. If requested (by flags), the event handler can 
then 
either compute the out-state (\STAGE{out_state_started} stage), or else revert 
to the \STAGE{event_time_started} stage. 

The strategy for suspending an event handler or for having it 
start an out-state computation (before the request) can be adjusted to the 
availability of physical processors on the multi-processor machine. However, in 
\JFVS, the communication \via pipes presents a computational bottleneck.

\subsection{Event handlers}
\label{sec:ArchitectureEventHandlers}

\begin{figure}[htb]
\begin{center}
\includegraphics{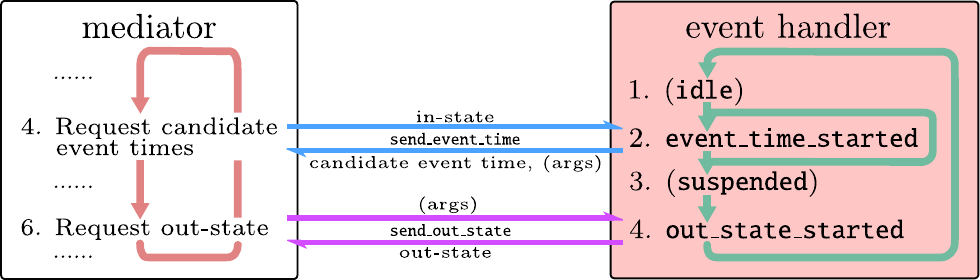}
\end{center}
\caption{Basic stages of event handlers for factors and pseudo-factors (stages 
1 and 3 relevant for the multi-process mediator only). In the 
\protect \STAGE{idle} and 
\protect \STAGE{suspended} stages, the event 
handler is halted (\via flags 
controlled by the multi-process mediator), thus liberating resources for other 
candidate-event-time requests. With the multi-process mediator, 
candidate 
out-states may be computed before the out-state request arrives.}
\label{fig:EventHandlerSingle}
\end{figure}

Event handlers (instances of a number of classes that inherit from the abstract 
\CLASS{EventHandler} class) provide the \METHOD{send_event_time} and 
\METHOD{send_out_state} methods that return candidate events. These candidate 
events either become events of a factor or pseudo-factor or they will be be 
trashed.\footnote{A candidate event time may stem from a  bounding potential, 
and not be confirmed for the factor potential. In \JFVS,
unconfirmed and confirmed events are treated alike.}

When realizing a factor or a pseudo-factor, event handlers receive the in-state 
as an argument of the \METHOD{send_event_time} method. The 
\METHOD{send_out_state} method then takes no argument. In contrast, event 
handlers that realize a set of factors or pseudo-factors request candidate event 
times without first specifying the complete in-state, because the element of the 
set that triggers the event is  yet unknown at the event-time request (see 
\sect{sec:EventHandlersSampling} for examples of event handlers that realize 
sets of factors). The \METHOD{send_event_time} method then takes the part of the 
in-state which is necessary to calculate the candidate event time. Also, it may 
return supplementary arguments together with the candidate event time, which is  
used by the mediator to construct the full in-state. The in-state is then 
an argument of the \METHOD{send_out_state} method, as it was not sent earlier. 

In \JFVS, each run requires a start-of-run event handler (an instance of a 
class that inherits from the abstract \CLASS{StartOfRunEventHandler}  class), 
and it 
cannot terminate properly without an end-of-run event handler. 
\sect{sec:EventHandlers} discusses several event-handler classes that are 
provided.

\subsection{State handler}
\label{sec:ArchitectureStateHandler}

The state handler (an instance of a class that inherits from the abstract 
\CLASS{StateHandler} class) is the sole separate element of \JF to access 
the global state. In \JFVS, the global physical state (all positions of \pps 
and 
\cpps) is contained in an instance of the \CLASS{TreePhysicalState} class 
represented as a tree consisting of nodes (each node corresponds to a 
\OBJECT{Node} object). Each node contains a particle (a \CLASS{Particle} 
object) 
which holds a time-sliced position. In \JFVS, each leaf node may in addition 
have charges as a Python dictionary mapping the name of the charge onto its 
value.

\begin{figure}[htb]
\begin{center}
\includegraphics{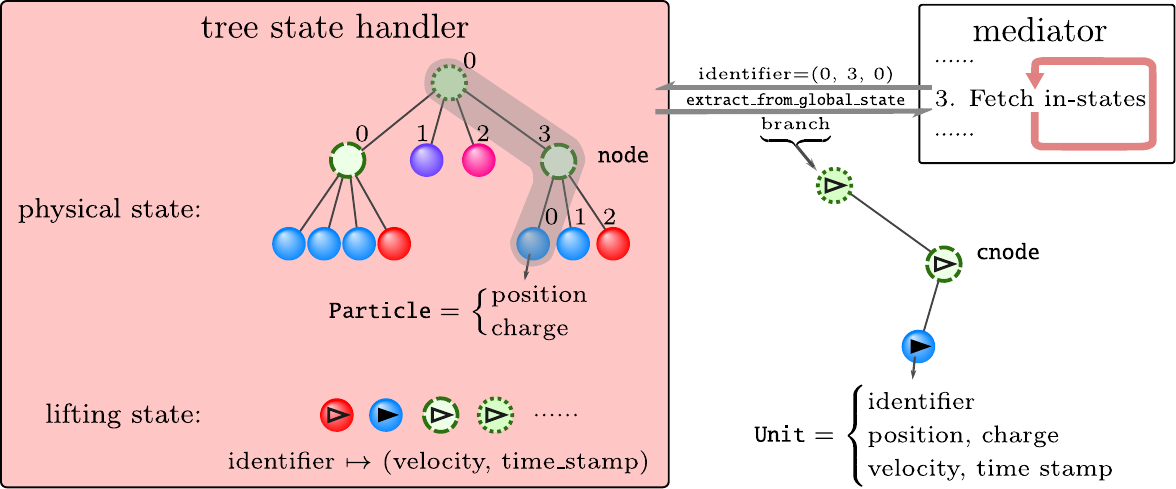}
\end{center}
\caption{Inner storage of the tree state
handler and example of its \protect \METHOD{extract_from_global_state} method, 
applied to the 
global state of \subfig{fig:CompositeParticleCells}{b}.}
\label{fig:StateHandler}
\end{figure}

Each tree is specified through its root node. Root nodes can be iterated over 
(in \JFVS, they are members of a list). Each node is connected to its parent 
and 
its children, which can also be iterated over. In \JFVS, the children are again 
members of a list. These lists imply unique identifiers of nodes and their 
particles as tuples. The first entry of the tuple gives a node's root node list 
index, followed by the indices on lower levels down to  the node itself (see 
\fig{fig:CompositeParticleCells}). 

The global lifting state is stored in \JFVS in a Python dictionary mapping the 
implicit particle identifier onto its time stamp and its velocity vector. This 
information is contained in an instance of the \CLASS{TreeLiftingState} 
class. Both the physical and lifting states are combined in the 
\CLASS{TreeStateHandler} which implements all methods of a state handler.

To communicate with other elements of the \JFA (such as the event handlers and 
the activator) \via the mediator, the state handler combines the information of 
the global physical and the global lifting state into units (that is, temporary 
\OBJECT{Unit} objects, see \fig{fig:StateHandler}). A given physical-state and 
lifting-state information for a node in the state handler is mirrored (that is 
copied) to a unit 
containing its implicit identifier, position, 
charge, velocity and time stamp. All other elements can access, modify, and 
return units. This provides a common packaging format across \JF. The 
explicit identifier of a unit  allows the program to integrate changed units 
into the  state handler's global state.

In the tree state handler of \JFVS, the local tree structure of nodes can be 
extracted into a branch of cnodes, that is, nodes containing 
units.\footnote{The 
distinction between particles and units, as well as between nodes and cnodes 
stresses that the state handler can only be accessed by the mediator, although 
information on the physical and the lifting state must of course travel 
throughout the application.} Each event handler only requires the global state 
reduced to a single factor 
in order to determine candidate event times and out-states. As a design 
principle in \JFVS, the event handlers keep the time-slicing of \cpps and its 
\pps consistent. Information sent to event handlers \via the mediator 
is therefore structured as branches, that is the information of a node with its 
ancestors and descendants. The state handler's 
\METHOD{extract_from_global_state} method 
creates a branch for a given identifier of a particle by constructing a 
temporary copy of the immutable node structure of the state handler using 
cnodes. 
Out-states of events in the form of branches can be 
committed to the global state using the \METHOD{insert_into_global_state} 
method.

The \METHOD{extract_active_global_state} method, 
the first of two additional methods provided by the state handler, 
extracts the part of the global 
state which appears in the global lifting state. The tree state handler 
constructs the minimal number of branches, where each node contains an active 
unit, so that all implicit identifiers appearing in the global lifting state 
are 
represented. 
The activator may then determine the factors which are to be 
activated. The method is also 
used to time-slice the entire global state (see 
\sect{sec:EventHandlersSampling}). Second, the 
\METHOD{extract_global_state} method extracts the full global state. (For the 
tree state handler of  \JFVS, this corresponds to  a branch 
for each root node.) This method 
does not copy the positions and velocities.

In \JFVS, the global physical state is initialized \via the input handler 
within the input--output handler (see \sect{sec:ArchitectureInputOutput}). The 
initial lifting state, however, is set \via the out-state of the 
start-of-run event handler, which is committed to the global state at the 
beginning of the program (see \sect{sec:EventHandlersSampling}). This means 
that, in \JFVS, the lifting state cannot be initialized from a file.

\subsection{Activator}
\label{sec:ArchitectureActivator}

The activator, a separate element of the \JF application, is an instance of a 
class that inherits from the abstract \CLASS{Activator} class. At the beginning 
of each leg, the activator provides to the mediator the new event handlers 
which are to be run, using the \METHOD{get_event_handlers_to_run} method. 
(As required by the mediator design pattern, no data flows directly between the 
activator and the event handlers, although it initially 
obtains their references, and subsequently manages them.) The activator also 
returns associated in-state identifiers of particles within the global state. 
The extracted parts of the global state of these are needed by the event 
handlers to compute their 
candidate event time (the identifier may be \pythoninline{None} if no 
information is needed). Finally, it readies for the mediator a list of 
trashable 
candidate events at the end of each leg in the \METHOD{get_trashable_events} 
method, once the mediator has committed the preceding event to the global state 
\via 
the state handler.

\begin{figure}[htb]
\begin{center}
\includegraphics{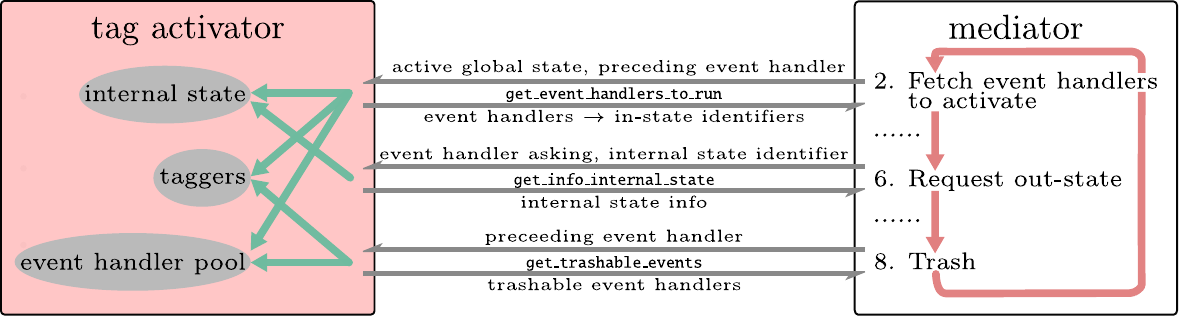}
\end{center}
\caption{Tag activator, and its complex interaction with the mediator. It 
readies event handlers and in-state identifiers, provides internal-state 
information for an out-state request, and identifies the trashable 
candidate events, as a function of the preceding event.}
\label{fig:Activator}
\end{figure}

In \JFVS, the activator is an instance of the \CLASS{TagActivator} class (that 
inherits from the \CLASS{Activator} class). The tag activator's operations 
depend on the interdependence of tags of event handlers and their events. Event 
handlers receive their tag by instances of classes located in the 
activator and derived from the abstract 
\CLASS{Tagger} class that are called \quot{taggers}.

A tagger centralizes common operations 
for identically tagged event handlers (see \fig{fig:Activator}). On 
initialization, the tagger receives 
its tag (a string-valued \VARIABLE{tag} attribute) and an event handler (that 
is, a single instance), of which it creates as many identical event-handler 
copies as needed (using the Python \pythoninline{deepcopy} method). Each tagger 
provides a \METHOD{yield_identifiers_send_event_time} method which generates 
in-state identifiers based on the branches containing independent active units 
(this means 
that the taggers are implemented especially for the \CLASS{TreeStateHandler}, 
the \CLASS{TagActivator} however is not restricted to this since it just 
transmits the extracted active global state). These in-states are passed (after 
extracting the part of the global state related to the identifiers from the 
state handler) to the \METHOD{send_event_time} method of the tagger's event 
handlers. The number of event handlers inside a tagger should meet the maximum 
number of events with the given tag simultaneously in the scheduler. In this 
paper, event handlers (and their candidate events) are referred to by tags, 
although in \JF they do not have the \VARIABLE{tag} attribute of their taggers.

On initialization, a tagger also receives a list of tags for event handlers that 
it creates, as well as a list of tags for event handlers that need to be 
trashed. The tag activator converts this information of all taggers into its 
internal \pythoninline{_create_taggers} and \pythoninline{_trash_taggers} 
dictionaries. Additionally, the tag activator creates an internal dictionary 
mapping from 
an event handler onto the corresponding tagger 
(\pythoninline{_event_handler_tagger_dictionary}).

A call of the \METHOD{get_event_handlers_to_run} method is accompanied by the 
event handler which created the preceding event
and by the 
extracted active global state. The event handler is first mapped onto its 
tagger. The taggers returned by the \pythoninline{_create_taggers} dictionary 
then generate the in-states identifiers, which are returned together with the 
corresponding event handlers (in a dictionary). For the initial call of the 
\METHOD{get_event_handlers_to_run} method no information on the preceding event 
handler can be provided. This is solved by initially returning the start-of-run 
event handler. Similarly the \pythoninline{_trash_taggers} dictionary is used on 
each call of \METHOD{get_trashable_events}. The corresponding event handlers are 
then also liberated, meaning that the activator can return them in the next call 
of the \METHOD{get_event_handlers_to_run} method.\footnote{The action of the 
\pythoninline{_create_taggers} and \pythoninline{_trash_traggers} dictionaries 
can be overruled with the concept of activated and deactivated taggers. Event 
handlers out of deactivated taggers are not returned to the mediator.} For this, 
the activator internally splits the pool of all event handlers of a given 
tag internally into those with a scheduled candidate event and the ones that are 
available to take on new candidate events.

The activator also maintains the internal state. In \JFVS, the internal state 
consists in cell-occupancy systems. Therefore, the internal state  is an 
instance of a class that inherits from the \CLASS{CellOccupancy} class, which 
itself inherits from the abstract \CLASS{InternalState} class. Taggers may refer 
to internal-state information to determine the in-states of their event 
handlers. The cell-occupancy system does not double up on the information 
available in the state handler. It keeps track of the identifier of a particle 
(which may correspond to  a \pp or a  \cpp), but does not store or copy the 
particle itself (see \sect{sec:SystemCell}). The mediator 
can access the internal state \via the \METHOD{get_info_internal_state} method 
(see \fig{fig:Activator}). To acquire consistency between the global state and 
the internal state (and between a particle and its associated unit), a 
pseudo-factor triggers an event for each active unit tracked by the 
cell-occupancy system that crosses a cell boundary (see 
\subfig{fig:FactorPseudoFactor}{b}). The internal state is updated in each call 
of the \METHOD{get_event_handlers_to_run} method.

\subsection{Scheduler}
\label{sec:ArchitectureScheduler}

The scheduler is an instance of a class inheriting from the abstract 
\CLASS{Scheduler} class. It keeps track of the candidate events and their 
associated event-handler references. Its \METHOD{get_succeeding_event} method 
selects among the candidate events the one with the smallest candidate event 
time, and it returns the reference of the corresponding event handler. Its 
\METHOD{push_event} method receives a new candidate event time and event-handler 
reference. Its \METHOD{trash_event} method eliminates a candidate event, based 
on the reference of its event handler. In \JFVS, the scheduler is an instance of 
the \CLASS{HeapScheduler} class. It implements a priority queue through the 
Python \MODULE{heapq} module.

\subsection{Input--output handler}
\label{sec:ArchitectureInputOutput}

The input--output handler is an instance of the \CLASS{InputOutputHandler} 
class. The input--output handler connects the \JFA to the outside world, and it 
is accessible by the mediator. The input--output handler breaks up into one 
input handler (an instance of a class that inherits from the abstract 
\CLASS{InputHandler} class) and a possibly empty list of output handlers 
(instances of classes that inherit from the abstract \CLASS{OutputHandler} 
class). These are accessed by the mediator only \via the input--output handler. 
Output handlers can also perform significant calculations.

The input handler enters the initial global physical state into the application. 
\JFVS provides an input handler that enters protein-data-bank formatted data 
(\PATH{.pdb} files) as well as an input handler which samples a random initial 
state. The initial state (constructed as a tree for the case of the tree state 
handler) is returned when calling the \METHOD{read} method of the input--output 
handler, which calls the \METHOD{read} method of the input handler.

The output handlers serve many purposes, from the output in \PATH{.pdb} files to 
the sampling of correlation functions and other observables, to a dump of the 
entire run. They obtain their arguments (for example the entire 
global state) \via its \METHOD{write} method. 
The \METHOD{write} method of the input--output handler receives
the desired output handler as an additional argument through
the mediating methods of specific event handlers.
These are triggered for example after a sampling or an
end-of-run event. 
The corresponding event handlers
are initialized with the name of their output handlers. 

\section{\JF event-handler classes}
\label{sec:EventHandlers}

Event handler classes differ in how they provide the \METHOD{send_event_time} 
and \METHOD{send_out_state} methods. Event handlers split into those that 
realize factors and sets of factors and those that realize pseudo-factors and 
sets of pseudo-factors. The first are required by ECMC while the second permit 
\JF to represent the entire run in terms of events. 

\subsection{Event handlers for factors or
sets of factors}
\label{sec:EventHandlerFactors}
Event handlers that realize a factor $M$, or a set  of factors are implemented 
in different ways depending on the analytic properties of the factor potential 
$U_M$ and on the number of independent active units. 

\subsubsection{Invertible-potential event handlers}
\label{sec:EventHandlerFactorsInv}

In \JF, an invertible factor potential $U_M$ (an instance of a class that 
inherits from the abstract \CLASS{InvertiblePotential} class) has its event rate 
integrated in closed form along a straight-line trajectory (see 
\fig{fig:EventFlow}). The sampled cumulative event rate ($U_M^+$ in~\cite[eq. 
(45)]{Faulkner2018}) provides the \METHOD{displacement} method. Together with 
the time stamp and the velocity of the active unit, this determines the 
candidate event time. In \JFVS, the two-leaf-unit  event handler (an instance of 
the \CLASS{TwoLeafUnitEventHandler} class) is characterized by two independent 
units at the leaf level. It realizes a two-particle factor with an invertible 
factor potential. The in-state (an argument of the \METHOD{send_event_time} 
method) is stored internally, and it remains available for the subsequent call 
of the \METHOD{send_out_state} method. Because of the two independent units, the 
lifting simply consists in these two units switching their velocities (using 
the 
internal \METHOD{_exchange_velocity} method) and keeping the velocities of all 
induced units consistent.

\subsubsection{Event handlers for factors with bounding potential}
\label{sec:EventHandlerFactorsBound}

For a factor potential $U_M$ that is not inverted (by choice or by necessity 
because it is non-invertible), the cumulative event rate $U_M^+$ is unavailable 
(or not used) and so is its \METHOD{displacement} method. Only 
the \METHOD{derivative} method is used. To realize such a factor without an 
inverted factor potential, an event handler then uses the \METHOD{displacement} 
method of an associated bounding potential whose event rate at least equals that 
of $U_M$ and that is itself invertible. A non-inverted $U_M$ may be associated 
with more than one bounding potential, each corresponding to a different event 
handler (the molecular Coulomb factors in \sect{sec:CookbookDipoles} associate 
the Coulomb factor potential in the same run with different bounding 
potentials). In \JFVS, a number of event handlers are instances of classes that 
inherit from the \CLASS{EventHandlerWithBoundingPotential} class, and that 
realize factors with bounding potentials. Each of these event handlers 
translates the sampled displacement of the bounding potential into a candidate 
event time. On an out-state request (\via the \METHOD{send_out_state} method), 
the event handler confirms the event with probability that is given by the ratio 
of the event rates of the factor potential and the bounding potential. The 
out-state consists of independent units together with their branches of induced 
units. For two independent units, the lifting limits itself to the application 
of a local \METHOD{_exchange_velocity} method, which exchanges independent-unit 
velocities and enforces velocities for the induced units. For more independent 
units, the out-state calculation requires a lifting. For an unconfirmed event, 
no lifting takes place. In \JFVS, confirmed and unconfirmed event have 
time-sliced out-states. Inefficient treatment of unconfirmed events is the 
main limitation of this version of the application.

A special case of a bounding potential is the cell-based bounding potential 
which features piecewise cell-bounded event rates. The two independent units are 
localized within their respective cells, and the bounding potential's rate is 
for all positions of the units larger than the factor potential event rate. In 
\JFVS, the constant cell-bounded event rate is determined for all pairs of cells 
on initialization (see \sect{sec:SystemPotentialsCellBounding}). The resulting 
displacement may move the independent active unit outside its cell. The proposed 
candidate event will then however be preempted by a cell-boundary event and then 
trashed (see \sect{sec:EventHandlersCellBoundary}).

\subsubsection{Cell-veto event handlers}
\label{sec:EventHandlersPairCellVeto}

Cell-veto event handlers  (instances of a number of classes that inherit from 
the abstract \CLASS{CellVetoEventHandler} class) realize sets of factors, rather 
than a single factor. The factor in-states (for each element of the set) are
not transmitted with the 
candidate-event-time request. Instead, the branch of the independent active 
unit is an argument of the \METHOD{send_event_time} method. 
The sampled factor in-state is transmitted with the out-state request. 
The 
cell-veto event handler implements Walker's 
algorithm~\cite{Walker1977AnEfficientMethod} in order to sample one 
element in the set of factors in \bigO{1} operations. 

Cell-veto event handlers are instantiated with an estimator (see 
\sect{sec:SystemEstimator}). In addition, they obtain a cell system which is 
read in through its \METHOD{initialize} method (see 
\sect{sec:SystemGlobalModules}). The estimator provides upper limits for the 
event rate (in the given direction of motion) for the independent active unit 
anywhere in one specific cell (called \quot{zero-cell}, see 
\sect{sec:SystemCell}), and for a target unit in any other cell, 
except for a list of excluded cells. These upper limits can be translated 
from the zero-cell to any other active-unit cell, 
because of the homogeneity of the simulation box. 
In \JFVS, the cell systems for 
the cell-veto event handler can be on any level of the particles' tree 
representation (see \sect{sec:CookbookSPCCVCV}, where a molecule-cell system 
tracks individual water molecules on the root level, while a oxygen-cell system 
tracks only the leaf nodes corresponding to oxygens).

A Walker sampler is an instance of the \CLASS{Walker} class in the 
\PACKAGE{event_handler} package. It provides the total event rate 
(\pythoninline{total_rate}), that for a homogeneous periodic  system 
is a constant 
throughout a run. On a candidate-event-time request, a cell-veto event handler 
computes a displacement, but no longer through the \METHOD{displacement} method 
of a factor potential or a bounding potential, but simply as an exponential 
random number divided by the total event rate. (The particularly simple 
\METHOD{send_event_time} method of a cell-veto event handler is implemented in 
the abstract \CLASS{CellVetoEventHandler} class.) (see~\cite{Faulkner2018} for a 
full description). The Walker sampler's \METHOD{sample_cell} method samples the 
cell of the target unit in \bigO{1}. It is returned, together with the 
candidate event time, as an argument of the \METHOD{send_event_time} method. The 
out-state request is accompanied by the branch of the independent unit in the 
target cell, if it exists. Confirmation of events and, possibly, lifting are 
handled as in \sect{sec:EventHandlerFactorsBound}.

\subsection{Event handlers for pseudo-factors or sets of pseudo-factors}

The pseudo-factors of \JF unify the description of the ECMC time evolution 
entirely in terms of events. The distinction between event handlers that realize 
pseudo-factors and those that realize sets of pseudo-factors remains crucial. In 
the former, the factor in-state is known at the candidate-event-time request. It 
is transmitted at this moment and kept in the memory of the event handler for 
use at the out-state request. For a set of pseudo-factors, the factor in-state 
can either not be specified at the candidate-event-time request, or would 
require transmitting too much data (one in-state per element of the set). It is 
therefore transmitted later, with the out-state-request (see 
\fig{fig:EndOfChain}).

\subsubsection{Cell-boundary event handler}
\label{sec:EventHandlersCellBoundary}

In the presence of a cell-occupancy system, \JFVS preserves consistency between 
the tracked particles of the global physical state  and the corresponding units 
(which must belong to the same cell). This is enforced by a cell-boundary event 
handler, an instance of the \CLASS{CellBoundaryEventHandler} class. This event 
handler has a single independent unit and realizes a pseudo-factor with a single 
identifier. A cell-boundary event leads to the internal state to be updated (see 
\sect{sec:ArchitectureActivator}). 

On instantiation, a cell-boundary event handler receives a cell system. (Each 
cell-occupancy system requires one independent cell-boundary event handler.) A 
candidate-event-time request by the mediator is accompanied by the in-state 
contained in a single branch and a single unit on the level tracked by 
the cell-occupancy system. An out-state request is met with the 
cell-level-unit's position corresponding to the minimal position in the new 
cell.

\subsubsection{Event handlers for sampling, end-of-chain, start-of-run, 
end-of-run}
\label{sec:EventHandlersSampling}

Sampling event handlers are instances of classes that inherit from the abstract 
\CLASS{SamplingEventHandler} class. Sampling event handlers are expected to 
produce output (they inherit from the \CLASS{EventHandlerWithOutputHandler} 
class and are connected, on instantiation, with their own output handler which 
is used in the mediating method of this event handler). 
Several sampling event handlers may coexist in one run. Their output handler is 
responsible for computing physical observable at the sampling event time (see 
\sect{sec:ArchitectureInputOutput}). \JFVS implements sampling events as the 
time-slicing of all the active units. A sampling event handler thus realizes a 
set of single-unit pseudo-factors, and the in-state is not specified at the 
candidate-event-time request. In \JFVS, the candidate event times of the 
sampling event handler are equally spaced. The out-state request is 
accompanied by branches of all independent active units, which are then all 
time-sliced simultaneously. Sampling candidate events are normally trashed only 
by themselves and by an end-of-run event.  

\label{sec:EventHandlersEOC}
\begin{figure}[htb]
\begin{center}
\includegraphics{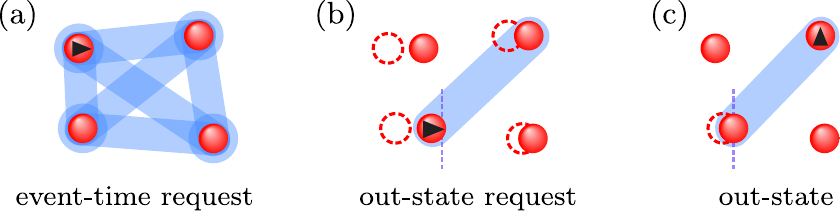}
\end{center}
\caption{Set of pseudo-factors realized by the end-of-chain event handler.
\subcap{a} Set of end-of-chain pair pseudo-factors for four \pps 
coupling the final active unit of the old chain and the beginning active unit 
of the new chain. 
\subcap{b} At the event time, the realized pseudo-factor with the incoming 
active unit and the outgoing unit is known. \subcap{c} A new event chain
is started. The outgoing active unit is shown. }
\label{fig:EndOfChain}
\end{figure}

End-of chain event handlers are instances of classes that inherit from the 
abstract \CLASS{EndOfChainEventHandler} class. They effectively stop one event 
chain and reinitialize a new one. This is often required for the entire run to 
be irreducible (see~\cite{Faulkner2018}). The end-of-chain event handler clearly 
realizes a set of pseudo-factors, rather than a single pseudo-factor (see 
\subfig{fig:EndOfChain}{a}). An end-of-chain event handler implements a method 
to sample a new direction of motion. In addition, it implements a method to 
determine a new chain length (that gives the time of the next end-of-chain 
event) and, finally, the identifiers of the next independent active cnodes.
For this, the end-of-chain event handler is aware of all the 
possible cnode identifiers (see \sect{sec:SystemGlobalModules}). 

On an event-time request, the end-of-chain event handler returns the next 
candidate event time (computed from the new chain length) and the identifier of 
the next independent active cnode. The  out-state request is accompanied by the 
current and the succeeding independent active units and their associated 
branches (see 
\subfig{fig:EndOfChain}{b}). For the out-state, the event handler determines the 
next direction of motion (see \subfig{fig:EndOfChain}{c}).

A start-of-run event handler (an instance of a class that inherits from the 
abstract \CLASS{StartOfRunEventHandler} class) is the sole event handler whose 
presence is 
required. The start-of-run event is the first one to be committed to 
the global state, because its candidate event time is set equal to the initial 
time of the run (usually zero) and because the activator will initially only 
activate the start-of-run event handler. The start-of-run event handler serves 
two purposes. First, it sets 
the initial lifting state. Second, the activator uses the start-of-run event 
handler as an entry point. Its tag (the \TAG{start_of_run} tag in the 
configuration files of \sect{sec:Cookbook}) is then used to determine the events 
that should be activated and created thereafter.

The end-of-run event handler (an instance of a class that inherits from the 
abstract \CLASS{EndOfRunEventHandler} class) terminates a run by raising an 
end-of-run exception and thus ends the mediator loop. An end-of-run event 
handler is usually connected, on instantiation, with its own output handler. In 
\JFVS, its \METHOD{send_event_time} method returns the total run-time, which 
transits from the configuration file. On the \METHOD{send_out_state} request, 
all active units are time-sliced. The end-of-run output handler may further 
process the global state which it receives \via the mediating method of the  
end-of-run event handler. 

\subsection{Event handlers for rigid motion of \cpps, mode switching}
\label{sec:EventHandlersCompositeParticleMotion}

The event handlers of \JFVS are generally suited for the rigid motion of \cpps 
(root mode), that is, for independent non-leaf-node units (as implemented in 
\sect{sec:CookbookDipolesLeavesRoots}). This is possible because all event 
handlers keep the branches of independent units consistent. As the subtree-node 
units of an independent-unit node move rigidly, the displacement is not 
irreducible. Mode switching into leaf mode (with single active leaf units) 
then 
becomes a necessity in order to have all factors be considered during
one run and to assure the irreversibility of the 
implemented 
algorithm. In \JFVS, the corresponding event handlers are instances of 
the \CLASS{RootLeafUnitActiveSwitcher} class.  On instantiation, they are 
specified to switch either from leaf mode to root mode or vice versa.

These event handlers resemble the end-of-chain event handler, but only one of 
them is active at any given time. They provide a  method to sample the new 
candidate event time based on the time stamp of the active independent unit at 
the time of its activation. An out-state request from one of these event 
handlers is accompanied by the entire tree of the current independent active 
unit of one mode and met with the tree of the  independent active unit on the 
alternate mode.

\section{\JF run specifications and tools}
\label{sec:System}

The \JF application relies on a user interface to select the physical system 
that is considered, and to fully specify the algorithm used to simulate 
it. Inside the application, some of these choices are made available to all 
modules (rather than having to be communicated repeatedly by the mediator). The 
application also relies on a number of tools that provide key features to many 
of its parts. 

\subsection{Configuration files, logging}
\label{sec:SystemUserInterface}

The user interface for each run of the \JFA  consists in a configuration file 
that is an argument of the executable \SCRIPT{run.py} 
script.\footnote{Configuration files follow the INI-file format and, in \JF, 
feature the
extension \INI{.ini}.} It specifies the 
physical and algorithmic parameters (temperature, system shape and size, 
dimension, type of \pps and \cpps, and also factors, 
factor potentials, lifting schemes, total run time, sampling frequency, etc). 

A configuration file is composed of sections that each correspond to a class 
requiring input parameters. The  \SECTION{Run} 
section specifies the mediator and the setting. The ensuing sections choose the 
parameters in the \METHOD{__init__} methods of the mediator and of the setting.
Each section contains 
pairs of properties 
and values. The property corresponds to the name of the argument in the 
\METHOD{__init__} method of the given class, and its value provides the 
argument 
(see \fig{fig:ConfigFileIni}). The content of the configuration file is parsed 
by the \MODULE{configparser} module and passed to the \JF factory (located in 
the 
\METHOD{base.factory} module) in \SCRIPT{run.py}. Standard Python naming 
conventions are respected in the classes built by the \JF factory, which 
implies the 
naming conventions in the configuration file (see 
\sect{sec:LicGitHubPythonPython} for details). Within the configuration file, 
sections can be written in any order, but their explicit nesting is not allowed. 
The nestedness is however implicit in the structure of the configuration file.

The \JFA returns all output \via files under the control of output 
handlers. 
Run-time information is logged
(the Python \MODULE{logging} module is used). Logged information can range from 
identification of CPUs to the initialization 
information of classes, run-time information, etc. Logging output (to standard 
output or to a file) can take place on a variety of levels from 
\LOGLEVEL{DEBUG} 
to \LOGLEVEL{INFO} to \LOGLEVEL{WARNING} that are controlled through 
arguments of \SCRIPT{run.py}. An identification hash of the run is part of the 
logging output. 
It also 
tags all the output files so that input, output and log 
files are uniquely linked (the Python \MODULE{uuid} module is used).

\subsection{Globally used modules}
\label{sec:SystemGlobalModules}

\JFVS requires that all trees representing \cpps are identical and of height at 
most two. Furthermore, in the $NVT$ physical ensemble, the particle number, 
system size 
and temperature remain unchanged throughout each run. After initialization, as 
specified in the configuration file, these parameters are stored in the \JF\ 
\PACKAGE{setting} package and the modules therein, which may be imported by all 
other modules, which can then autonomously construct identifiers. Helper 
functions for periodic boundary conditions (if available) and for  the sampling 
of random positions are also accessible.

\JFVS implements hypercubic and hypercuboid setting modules. Both settings 
define the inverse temperature and also the attributes of all possible particle 
identifiers, which are broadcast directly by the \PACKAGE{setting} package. In 
contrast, the parameters of the physical system are accessed only using the 
modules of the specific setting (for example the 
\MODULE{setting.hypercubic_setting} module).\footnote{Attributes in the 
\PACKAGE{setting} package are copied to the modules for convenience.} The 
\PACKAGE{setting} package and its modules are initialized by classes which 
inherit from the abstract \CLASS{Setting} class. The \CLASS{HypercuboidSetting} 
class defines only the hypercuboid setting, the \CLASS{HypercubicSetting} class, 
however, sets up both the \MODULE{hypercubic_setting} and the 
\MODULE{hypercuboid_setting} modules together with the \PACKAGE{setting} 
package. This allows modules that are specifically implemented for a hypercuboid 
setting to be used with the hypercubic setting.

Each setting can implement periodic boundaries, by inheriting from the abstract 
\CLASS{PeriodicBoundaries} class and by implementing its methods. Since many 
modules of \JF only rely on periodic boundaries but not on the specific setting, 
the \PACKAGE{setting} package gives also access to the initialized periodic 
boundary conditions. Similarly, a function to create a random position is 
broadcast by the setting package. All the configuration files in 
\sect{sec:Cookbook} are for a three-dimensional cubic simulation box, that is, 
use the 
hypercubic setting with \pythoninline{dimension = 3}.

Additional useful modules are located in the \JF\ \PACKAGE{base} package. The 
abstract \CLASS{Initializer} class located in the \MODULE{initializer} module 
enforces the implementation of an \METHOD{initialize} method. This method must 
be called before other public methods of the inheriting class. The 
\MODULE{strings} module provides functions to translate strings from snake to 
camel case and vice versa, as well as to translate a package path into a 
directory path. Helper functions for vectors, such as calculating the norm or 
the dot product, are located in the \MODULE{vectors} module.

\subsection{Cell systems and cell-occupancy systems} 
\label{sec:SystemCell} 

\begin{figure}[htb]
\begin{center}
\includegraphics{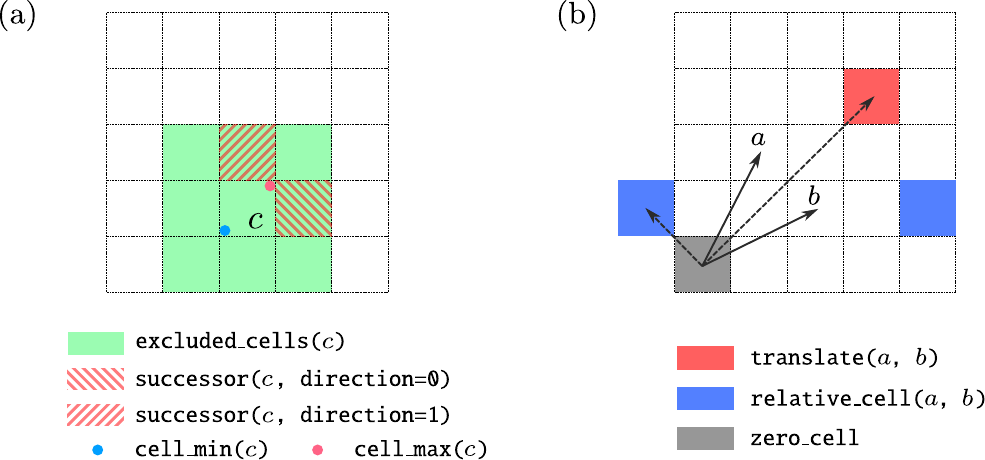}
\end{center}
\caption{Cell methods. \subcap{a} \protect \METHOD{excluded_cells}, \protect 
\METHOD{successor}, \protect \METHOD{cell_min} and \protect \METHOD{cell_max} 
methods required by the abstract \protect \CLASS{Cells} class.
Horizontal and vertical directions are indexed as $0$ and $1$, 
respectively. \subcap{b} \protect \METHOD{translate} and \protect 
\METHOD{relative_cell} methods (illustrated by vectors) required
by the \protect \CLASS{PeriodicCells} class, in 
addition to the methods of the \protect \CLASS{Cells} class.
Periodic boundary conditions are required, and the two blue cells are identical.
The periodic-cell system's origin is given by the \protect 
\pythoninline{zero_cell} 
property.}
\label{fig:CellMethods}
\end{figure}

\begin{figure}[htb]
\begin{center}
\includegraphics{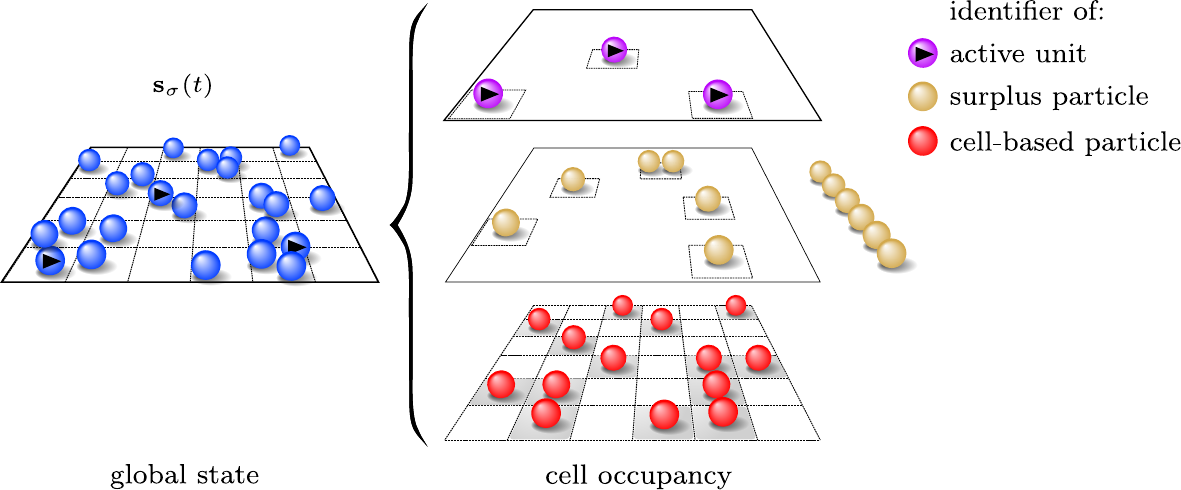} 
\end{center}
\caption{Cell-occupancy system, an internal state of the activator, with active 
units accounted for differently from surplus and cell-based particles. Only a 
fixed number of cell-based particle identifiers are allowed per cell (here one 
per cell). Surplus-particle identifiers may be iterated over from the outside of 
the cell-occupancy system  with a  \protect \METHOD{yield_surplus} method. In 
\JFVS, surplus particles form an internal dictionary mapping the cell onto the 
particle identifier. }
\label{fig:CellSystem3D}
\end{figure}

A cell-occupancy system is an instance of  a class that inherits from the 
abstract \CLASS{CellOccupancy} class, located in the activator. Any 
cell-occupancy system is associated with a cell system, itself an instance of a 
class that inherits from the abstract \CLASS{Cells} class.

In \JFVS, the cell system consists in a regular grid of cells that are referred 
to through their indices. Cells can be iterated over with the 
\METHOD{yield_cells} method. For a given cell, the excluded cells are accessed
by the \METHOD{excluded_cells} method, the successor cell in a suitably 
defined direction by the \METHOD{successor} method and the lower and upper bound 
position in each direction through the \METHOD{cell_min} and 
\METHOD{cell_max} methods (see \subfig{fig:CellMethods}{a}). Finally, the 
\METHOD{position_to_cell} method returns the cell for a given position. Cell 
systems with  periodic boundary conditions are described as periodic cell 
systems (instances of classes that inherit from the abstract 
\CLASS{PeriodicCells} class, which itself inherits from the
\CLASS{Cells} class). Their \pythoninline{zero_cell} property
corresponds to the cell located at the origin. Their 
\METHOD{relative_cell} method receives a cell and a reference cell, and 
establishes equivalence between  the relative and the zero-cell. The inverse to 
this is the \METHOD{translate} method (see \subfig{fig:CellMethods}{b}). 

A cell-occupancy system (which is located in the activator) associates the 
identifiers  of cell-based particles and of surplus particles with a cell. It 
also stores active cells, that is, cells that contain an active unit (see 
\fig{fig:CellSystem3D}). Cell-based and surplus particles in the state handler 
correspond to units with zero velocity, so that there is no real distinction 
between units and particles for them. The cell-occupancy system inherits from 
the abstract \CLASS{InternalState} class and therefore provides 
\METHOD{__getitem__} and \METHOD{update} methods. The former returns a particle 
identifier based on a cell, whereas the latter updates the cell 
occupancies based on the currently active units. This keeps the internal state 
consistent with the global state. Moreover the cell-occupancy may iterate over 
surplus particle identifiers \via the \METHOD{yield_surplus} method. The active 
cells and the corresponding identifiers of the active units are generated using 
the \METHOD{yield_active_cells} method (see \fig{fig:CellSystem3D}).

\JFVS implements the \CLASS{SingleActiveCellOccupancy} class which features only 
a single active cell and which keeps the active unit identifier among its 
private attributes. The cell-based particle identifiers are stored in an 
internal \pythoninline{_occupant} list, and surplus-particle identifiers are 
stored in an internal  \pythoninline{_surplus} dictionary mapping the cell 
indices onto the surplus-particle identifiers.

The stored cell-occupancy system can address different levels of composite 
particles: one cell-occupancy system may track particles (and units) associated 
to root nodes, and another one particles that go with leaf nodes. This is set on 
initialization \via the \VARIABLE{cell\_level} property which equals the length 
of the particle identifier tuple. The concerned  cell system is itself set on 
initialization. An indicator charge allows one to select specific particles on a 
given level for tracking.

A single run can feature several internal states stored within the activator. 
These instances may rely on different cell-occupancy systems and cell systems. 
For consistency between  internal states and the global state, each 
cell-occupancy system requires its own cell-boundary event handler.

\subsection{Inter-particle potentials and bounding potentials}
\label{sec:SystemPotentials}

In \JF, potentials play a dual role, as factor potentials $U_M$ to event 
handlers 
but also as bounding potentials for factor potentials $U_M$. Potentials are 
located in the  \JF\ \PACKAGE{potential} package. They inherit from the 
abstract \CLASS{Potential} class and provide a \METHOD{derivative} method. They 
may in addition inherit from the abstract \CLASS{InvertiblePotential} class, and 
must then provide a \METHOD{displacement} method. In \JFVS, derivatives and 
displacements are with respect to the 
positive change of the active unit along one of the coordinates  (indicated 
through the \pythoninline{direction}). For a potential $U(\rvec_{ij})$ and 
\pythoninline{direction} $=0$, the derivative is for example given by 
$\glc \partial /\partial x_i U(\rvec_{ij}) \grc$.

\subsubsection{Inverse-power-law potential, Lennard-Jones potential}
\label{sec:SystemPotentialsInverse}

The inverse-power-law potential (an instance of the 
\CLASS{InversePowerPotential} class that inherits from the abstract 
\CLASS{InvertiblePotential} class) concerns the 
separation vector $\rvec_{ij}=\rvec_{j}-\rvec_{i}$ 
(without periodic boundary conditions, in $d$-dimensional space)
between a unit $j$ 
and an active unit $i$ as
\begin{equation}
	U_{(\SET{i,j},\TYPE{inv})}(\rvec_{ij},c_i, c_j)=c_i c_j k
	\left(\frac{1}{|\rvec_{ij}|}\right)^{p}.
\label{equ:InversePowerLawPotential}
\end{equation}
Here, $k$ and $p>0$ correspond to the \pythoninline{prefactor} and 
\pythoninline{power} parameters set on initialization. The 
charges $c_i$ and $c_j$ are entered into the methods of the potential as 
parameters \pythoninline{charge_one} 
and \pythoninline{charge_two}. This allows one instance of the
\CLASS{InversePowerPotential} class to be used for different charges.
The \METHOD{derivative} method is straightforward, while the 
\METHOD{displacement} method distinguishes the repulsive ($c_i c_j k>0$) and 
the attractive ($c_i c_j k<0$) cases.

The Lennard-Jones potential (an instance of the 
\CLASS{LennardJonesPotential} class) implements the Lennard-Jones potential
\begin{equation}
U_{\FACTOR{\SET{i,j}}{\TYPE{\LJTYPE}  }}(\rvec_{ij}) = k_{\LJTYPE} \glc \glb
\frac{\sigma }{ |\rvec_{ij}|} \grb^{12}
- \glb \frac{\sigma }{ |\rvec_{ij}|} \grb ^6 \grc,
\label{equ:LJFACTORINTRO}
\end{equation}
where $\rvec_{ij} = \rvec_j - \rvec_i$ is the separation vector (without 
periodic boundary conditions, in $d$-dimensional space) between a unit $j$ and 
an active unit $i$. and 
where $k_{\LJTYPE}$ and $\sigma$ correspond to the parameters 
\pythoninline{prefactor} and \pythoninline{characteristic_length} set on 
instantiation. This Lennard-Jones potential provides a straightforward 
\METHOD{derivative} method. Its \METHOD{displacement} method relies on an 
algebraic inversion.

\subsubsection{Displaced-even-power-law potential}
\label{sec:SystemPotentialsDisplaced}

An instance of the 
\CLASS{DisplacedEvenPowerPotential} class that inherits from the abstract
\CLASS{InvertiblePotential} class, 
the displaced-even-power-law potential, concerns
the separation vector $\rvec_{ij}=\rvec_{j}-\rvec_{i}$ 
(without periodic boundary conditions, 
in $d$-dimensional space)
between a unit $j$ and an active unit $i$ 
\begin{equation}
U_{(\SET{i,j},\TYPE{depp})}(\rvec_{ij})=
k_\text{depp} \glb |\rvec_{ij}|-r_0\grb ^p,
\label{equ:DisplacedEvenPowerLaw}
\end{equation}
where $k_\text{depp} > 0$, $p \in \SET{2,4,6,\dots}$, and $r_0$, respectively, 
are 
the parameters \pythoninline{prefactor}, \pythoninline{power}, and 
\pythoninline{equilibrium_separation} parameters
set on instantiation. The 
\METHOD{derivative} and \METHOD{displacement} methods are provided 
analytically. 

\subsubsection{Merged-image Coulomb  potential and bounding potential}
\label{sec:SystemPotentialsCoulombMerged}
An instance of the 
\CLASS{MergedImageCoulombPotential} class that inherits from the abstract 
\CLASS{Potential} class, the merged-image Coulomb potential
is defined for a separation vector 
$\rvec_{ij}=\rvec_{j}-\rvec_{i}$ 
(with periodic boundary conditions, in three-dimensional space)
between a unit $j$ and an active unit $i$ as
\begin{equation}
U_{\rm C}(\rvec_{ij}, c_i, c_j)=\sum_{\nvec} 
c_i c_j / | \rvec_{ij} + \nvec \Lvec |,\quad \nvec\in \ZZ^3, 
\label{equ:CoulNaive}
\end{equation}
where  $\Lvec = (L_x, L_y, L_x)$ are the sides of the three-dimensional 
simulation box with periodic boundary conditions. 
The  conditionally convergent sum in \eq{equ:CoulNaive} can be consistently 
defined in terms of 
\quot{tin-foil} boundary conditions~\cite{deLeeuw1980-2}. 
It then yields an absolutely convergent sum, partly in real space and partly in 
Fourier space (see~\cite[Sect. IIIA]{Faulkner2018}), 
\begin{equation}
\psi(\rvec_{ij}, c_i, c_j) =c_i c_j\glc \sum_{\nvec \in \ZZ ^3} \frac{{\rm 
erfc} 
(\alpha |
\rvec_{ij}+\nvec L |)}{|
\rvec_{ij} + \nvec L |} \right.
 \left. \!\! + \frac{4 \pi }{L^3} \!\!\!\sum_{\qvec\ne (0,0,0)} \!\!\!\!\!\!
\frac{\expa{- \qvec^2 / (4\alpha^2)}}{\qvec^2}\cosb{\qvec\cdot \rvec_{ij}} \grc
,
\label{equ:EwaldCoulFunction}
\end{equation}
with $\alpha$ a tuning parameter and $\qvec=2 \pi \mvec/L$, $\mvec \in \ZZ^3$. 
\JFVS provides this class for a cubic simulation box, with parameters that are 
optimized to 
reach machine precision for its \pythoninline{derivative} method. Summations 
over $\nvec$ and $\mvec$ are taken within spherical cutoffs, namely for all 
$|\nvec|\le\pythoninline{position_cutoff}$ and 
$|\mvec|\le\pythoninline{fourier_cutoff}$ except that $\mvec=(0, 0, 0)$. (The 
potential in \eq{equ:EwaldCoulFunction} differs from the tin-foil Coulomb 
potential in a constant self-energy term that does not influence the 
derivatives.)

The merged-image Coulomb potential is not invertible. When it serves as a 
factor potential, bounding potentials provide the 
required 
\pythoninline{displacement} method. 
\JFVS provides a 
merged-image Coulomb bounding potential as
an instance of the \CLASS{InversePowerCoulombBoundingPotential} class, with
\begin{equation}
U_{\rm C, \text{Bounding}}=c_i c_j k_b/{|\rvec_{ij, 0}|}.
\label{equ:CoulBounding}
\end{equation}
Here, $\rvec_{ij, 0}$ is the minimum separation vector, that is, the vector 
between
$\rvec_i$ and the closest image of $\rvec_j$ under 
the periodic boundary conditions. (The merged-image Coulomb bounding 
potential thus involves no sum over periodic images.) The constant 
$k_b$ is chosen as
\begin{equation}
 k_b = \max_{\rvec\in [-L/2, L/2]^3} 
 \frac{|\rvec|^3}{x}\partpart{\psi(\rvec)}{x},
\end{equation}
so that the factor-potential event rate is bounded. A constant $k_b \gtrsim 
1.5836$ (the 
parameter \pythoninline{prefactor}) is appropriate for a cubic simulation box.
The merged-image Coulomb bounding potential is closely related to the 
inverse-power-law potential of \eq{equ:InversePowerLawPotential} with
$p=1$, although the restriction to the minimum separation vector makes that the 
latter cannot be used directly.

\subsubsection{Cell-based bounding potential}
\label{sec:SystemPotentialsCellBounding}

A cell-based bounding potential is an instance of a class that inherits from the 
abstract \CLASS{InvertiblePotential} class. It bounds the derivative of the 
factor potential inside certain cell regions by constants. These constants can 
be computed analytically on demand or even sampled using a 
separate Monte Carlo algorithm. On initialization, a cell-based bounding 
potential receives an estimator (see \sect{sec:SystemEstimator}). Also the 
information about the cell system is transmitted. Then, the cell-based bounding 
potential iterates over all pairs of cells (making use of periodic boundary 
conditions) and determines an upper and lower bound derivative for the factor 
units being in those cells for each possible 
direction of motion using the estimator. Here, the cell-based bounding potential 
is not applied to excluded cells, where the cell-bounded event rate diverges, is 
simply too large, or otherwise inappropriate.

The constant-derivative bound leads to a piecewise linear invertible bounding 
potential. The call of the \METHOD{displacement} method is accompanied by the 
direction of motion, the charge product, the sampled potential change and the 
cell separation. In \JFVS, any cell-based bounding potential requires a 
cell-boundary event handler, that detects when the displacement proposed by the 
\METHOD{displacement} method in fact takes place outside the cell for which it 
is computed. 

\subsubsection{Three-body bending potential}
\label{sec:SystemPotentialsBending}

The SPC/Fw water model of \sect{sec:CookbookSPC} includes a bending potential 
(an instance of the 
\CLASS{BendingPotential} class), which describes the 
fluctuations in the bond angle within each molecule. For the three 
units $i$, $j$, and $k$ within such a molecule in three-dimensional space (with 
$j$ being the oxygen), it 
is given by
\begin{equation}
U_{(\SET{i,j,k},\TYPE{\BENDINGTYPE})}(\rvec_{ij},\rvec_{jk})
=\frac{1}{2}k_b\Big(\phi_{\SET{i,j,k}}(\rvec_{ij},\rvec_{jk})-\phi_0\Big)^2.
\end{equation}
Here, $\phi_{\SET{i,j,k}}(\rvec_{ij},\rvec_{jk})$ denotes the internal angle 
between the two hydrogen--oxygen legs. The constants $k_b$ and $\phi_0$ are set 
on initialization of the potential (see~\cite{Faulkner2018}). The 
\METHOD{derivative} method is provided explicitly for this potential, which is 
however not invertible. 

In \JFVS, the associated bounding potential is constructed dynamically by an 
event handler\footnote{instance of the 
\CLASS{FixedSeparationsEventHandlerWithPiecewiseConstantBoundingPotential} class 
} which dynamically constructs a piecewise linear bounding potential. Here, the 
event handler speculates on a constant bounding event rate through its position 
between two subsequent time-sliced positions of the active unit:
$q_{\rm bounding}=\max\{q(\rvec), 
q(\rvec+\vvec \Delta t)\}+\const$ where $q(\rvec)$ is the potential derivative 
at $\rvec$.
The interval length $|\vvec \Delta t|$ and the constant offset are input from 
the configuration file. Fine-tuning provides an efficient bounding potential 
that does not under-estimate the event rate, yet limits the ratio of 
unconfirmed events. 

\subsection{Lifting schemes}
\label{sec:SystemLifting}

Event handlers with more than two independent units require a lifting scheme (an 
instance of a class that inherits from the \CLASS{Lifting} class). The event 
handler calls a method of the lifting scheme to compute its out-state. At first, 
the event handler prepares factor derivatives of relevant time-sliced units. 
The derivative table (see~\cite[Figs 2 and 10]{Faulkner2018}) is filled with 
unit identifiers, factor derivatives and activity information 
through its \METHOD{insert} method. 
Finally, the event handler calls the 
\METHOD{get_active_identifier} method that returns the identifier of the next 
independent active unit. The lifting scheme's \METHOD{reset} method deletes the 
derivative table. It is called before the first derivative is inserted.
\JFVS implements the ratio, inside-first and outside-first 
lifting schemes
for a single independent active unit (see~\cite[Sect. IV]{Faulkner2018}).

\subsection{Estimator}
\label{sec:SystemEstimator}

Estimators (instances of a class that inherits from the abstract 
\CLASS{Estimator} class) determine upper and lower bounds on the factor 
derivative in a single direction between a minimum and maximum corner of a 
hypercuboid for the possible separations. For this, they provide the 
\METHOD{derivative_bound} method. Both upper and lower bounds are useful when 
the potential can have either positive and negative  charge products (as 
happens for example for the merged-image Coulomb potential as a function of the 
two charges). In general, an estimator compares the factor derivatives for 
different separations in the hypercuboid to obtain the bounds. These are 
corrected by a prefactor and optionally by an empirical bound, which are set on 
instantiation (together with the factor potential).

\JFVS provides estimators which either regard regularly or randomly sampled 
separations within the hypercuboid. The inner-point and boundary-point 
estimators vary the separation evenly within the hypercuboid or on the edge of 
the hypercuboid, respectively. For these separations, the factor potential 
derivatives (optionally including charges) are compared. Two more estimators
consider the interaction between a charged active unit and two oppositely 
charged target units within a dipole. Here, the factor derivative is summed for 
the two possible active-target pairs. A Monte-Carlo estimator distributes both 
the separation and the 
dipole orientation randomly. The dipole-inner-point estimator 
varies the separations evenly but aligns the dipole orientation along the 
direction of the gradient of the factor derivative. The implemented estimators 
are appropriate for the cookbook examples of \sect{sec:Cookbook}, where the 
upper and lower bounds on the factor derivatives (and equivalently on the
event rates) must be computed for a small number of cell pairs only.

\section{\JF Cookbook} \label{sec:Cookbook} 
The configuration files\footnote{Configuration files  in the  
\PATH{src/config_files/2018_JCP_149_064113} directory tree are described in 
this section.}  in \JFVS introduce to the key features of the application by 
constructing runs for two charged \pps, for two interacting dipoles of charges, 
and for two interacting water molecules (using the SPC/Fw model). All 
configuration files are for a three-dimensional cubic simulation box with 
periodic boundary conditions, and they reproduce published 
data~\cite{Faulkner2018}. 

\begin{figure}
\begin{center}
\includegraphics{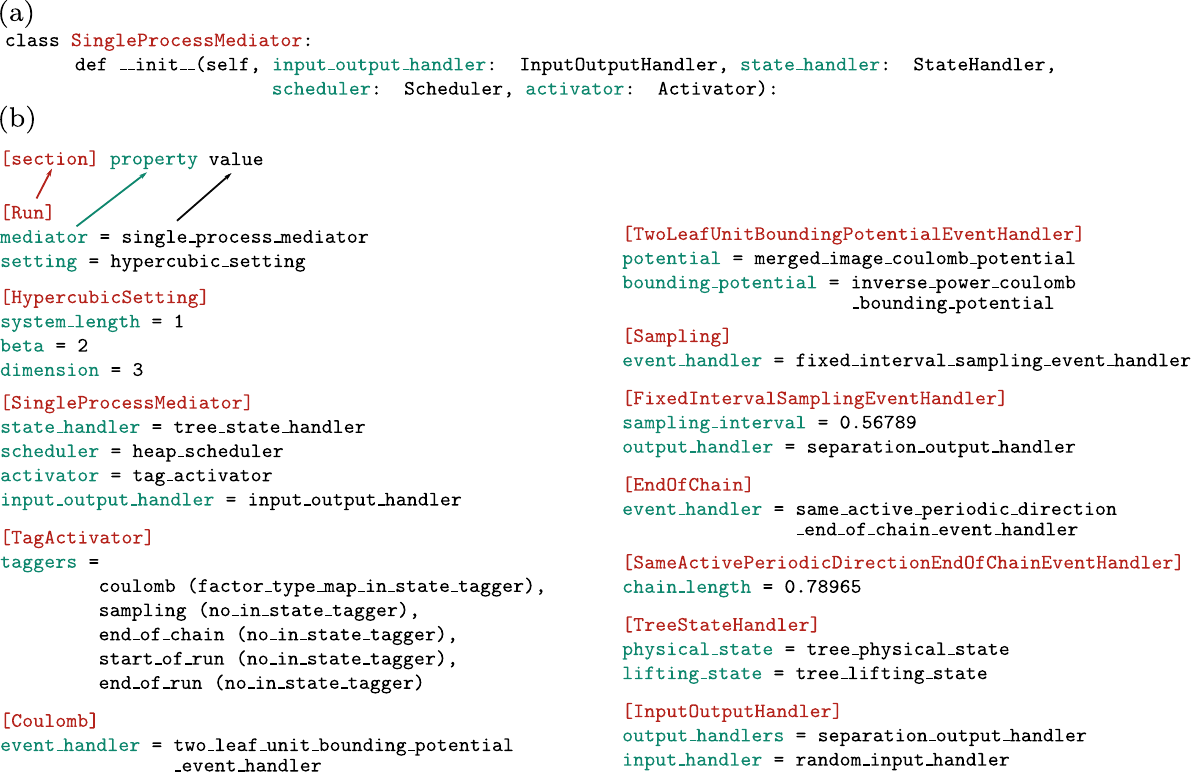}
\end{center}
\caption{Configuration file \protect \INI{coulomb_atoms/power_bounded.ini}. 
\subcap{a} A typical \protect \pythoninline{__init__} method of a \JF class. 
\subcap{b} Excerpts of the configuration file (some lines split for clarity). 
Sections with properties and values that correspond to the argument names in 
the 
 \protect \pythoninline{__init__} methods of \JF classes. }
\label{fig:ConfigFileIni}
\end{figure}

As specified in their 
\SECTION{Run} 
sections, 
the configuration files use a single-process mediator (an instance of  the 
\CLASS{SingleProcessMediator} class), and the \PACKAGE{setting} package is 
initialized by an instance of the \CLASS{HypercubicSetting} class
(see for example \fig{fig:ConfigFileIni}). All configuration files  in 
the directory use a heap scheduler (an instance of the \CLASS{HeapScheduler} 
class), a tree state handler (instance of the \CLASS{TreeStateHandler} 
class), as well as an tag activator (an instance of the \CLASS{TagActivator} 
class) in order to activate event handlers, trash candidate events and prepare 
in-states.

The \TAG{start_of_run}, \TAG{end_of_run}, \TAG{end_of_chain}, and 
\TAG{sampling} event handlers (that realize common pseudo-factors)
are implemented in largely analogous sections across all the 
configuration files, although their parent sections (that define the 
corresponding taggers) provide different tag lists for trashing and activation 
of event handlers. The corresponding tagger sections are presented in detail in 
\sect{sec:CookbookAtomsNaive}, and only briefly summarized thereafter.

\subsection{Interacting atoms}
\label{sec:CookbookAtoms}

The configuration files in the \pythoninline{coulomb_atoms} directory of \JFVS 
implement the ECMC sampling of the Boltzmann distribution for two identical 
charged \pps. They interact with the merged-image Coulomb pair potential and are 
described by a Coulomb pair factor. One of the two \pps is active, and it moves 
either in $+x$, in $+y$, or in $+z$ direction. Statistically equivalent output 
is obtained for the merged-image Coulomb pair potential (the factor potential) 
associated with the inverse-power bounding potential 
(\sect{sec:CookbookAtomsNaive}), or else with a cell-based bounding 
potential, either realized directly (\sect{sec:CookbookAtomsCell}), or through a 
cell-veto event handler (\sect{sec:CookbookAtomsVeto}). Although the 
configuration files use the language of \sect{sec:IntroductionGlobalInternal} 
for the representation of particles, all trees and branches are trivial, and 
each root node is also a leaf node.

\subsubsection{Atomic factors, inverse-power Coulomb bounding potential}
\label{sec:CookbookAtomsNaive}

\begin{figure}
\begin{center}
\includegraphics{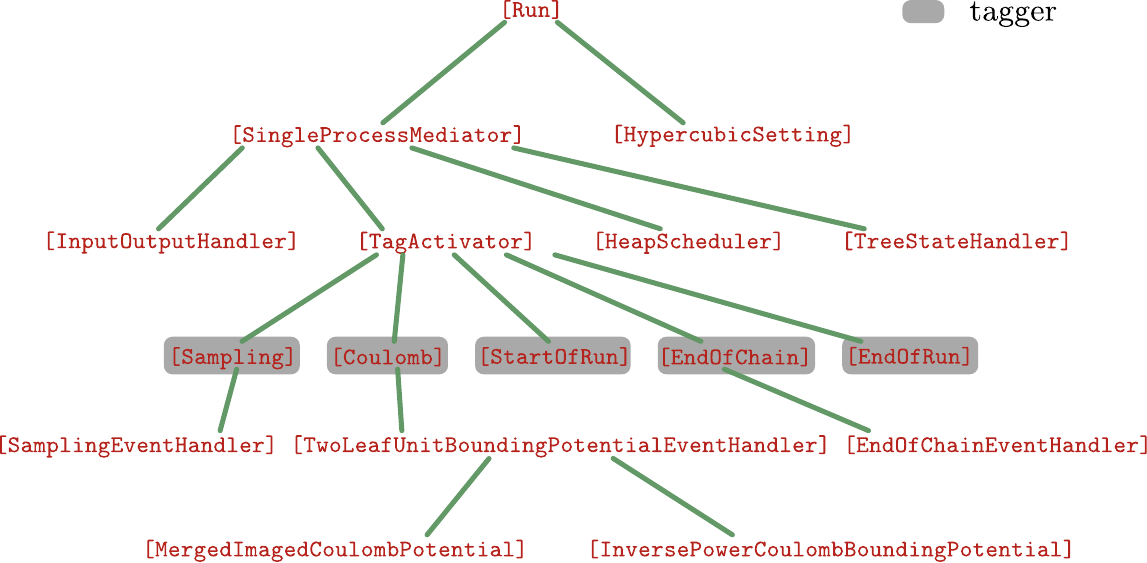}
\end{center}
\caption{Tree representation the sections in the configuration file \protect 
\INI{coulomb_atoms/power_bounded.ini}. Only part of the tree is shown and names 
of event handlers for sampling and end-of-chain are shortened. The children of 
the \protect \SECTION{TagActivator} section correspond to all the declared 
taggers, which point towards sections for their associated event-handler 
classes. }
\label{fig:ConfigFileTree}
\end{figure}

The configuration file \INI{coulomb_atoms/power_bounded.ini} implements a 
single 
Coulomb pair factor with the merged-image Coulomb factor potential 
that is associated with its
inverse-power Coulomb 
bounding potential. The same event handler realizes this factor for any 
separation of the \pps. The activator requires no internal state.

Although it would be feasible to directly implement (that is, hard-wire) all 
event 
handlers for this simple system,  the tag activator is used. All event handlers 
are thus accessed \via taggers that are listed, together with their tags, in 
the 
\SECTION{TagActivator} section (see \fig{fig:ConfigFileTree} for a tree 
representation of the sections). The \TAG{coulomb} tagger is an instance of the 
\CLASS{FactorTypeMapInStateTagger} class, indicating that its event 
handlers require a specific
in-state created from a pattern stored in a file indicated in the 
\SECTION{FactorTypeMaps} section. This pattern mirrors the factor index sets 
and 
factor types for a system with two root nodes (see 
\eq{equ:PotentialFactorized}). The entry \pythoninline{[0, 1], Coulomb} in this 
file  indicates 
that, for two \pps, a Coulomb potential would act between particles 
0 and 1. From this information, the tagger's 
\METHOD{yield_identifiers_send_event_time} method generates all the in-state 
identifier for any number of \pps.

The \SECTION{Coulomb} section specifies input for the \TAG{coulomb} tagger's 
tag 
lists (the \pythoninline{creates} list and the \pythoninline{trashes} list). 
Here, a \TAG{coulomb} event creates and trashes only \TAG{coulomb} candidate 
events (see the configuration file of \sect{sec:CookbookSPCAtomicInv} for 
different tag lists for the same \TAG{coulomb} event handlers).

The \SECTION{Coulomb} section further specifies that the \TAG{coulomb} event 
handler is an instance of the \CLASS{TwoLeafUnitBoundingPotentialEventHandler} 
class and that, for two \pps, only one \TAG{coulomb} event handler is needed. 
The corresponding section\footnote{The 
\SECTION{TwoLeafUnitBoundingPotentialEventHandler} section. The section name 
may 
be replaced by an alias to respect the tree structure of the configuration file 
(see \sect{sec:CookbookDipolesLeaves}). } specifies the factor potential to be 
an instance of the \CLASS{MergedImageCoulombPotential} class. It specifies the 
bounding potential as an instance of the 
\CLASS{InversePowerCoulombBoundingPotential} class. The \TAG{sampling}, 
\TAG{end_of_chain}, \TAG{start_of_run} and \TAG{end_of_run} taggers are all 
instances of the \CLASS{NoInStateTagger} class (their event handlers require no 
in-state), and also provide their event handlers and their tag lists, which are 
then transmitted to the tag activator. Each of these taggers' 
\METHOD{yield_identifiers_send_event_time} methods yields the in-state 
identifiers needed by the taggers' event handlers in order to realize 
corresponding factors or pseudo-factors.

\begin{figure}[htb]
\begin{center}
\includegraphics{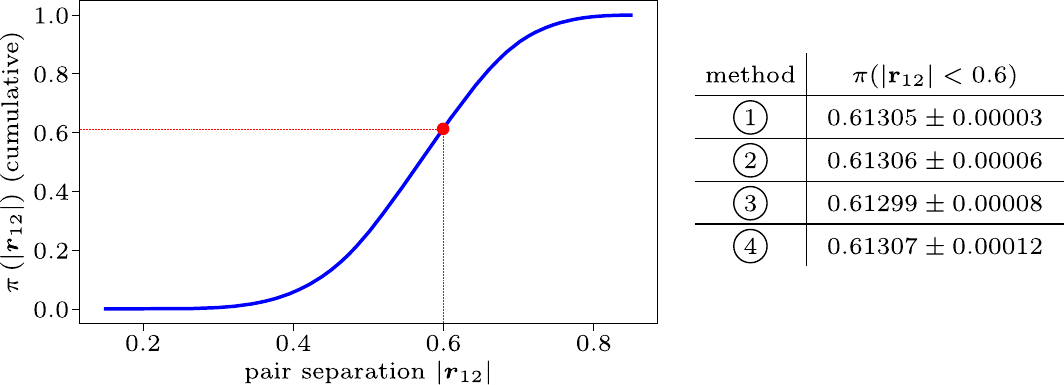}
\end{center}
\caption{
Cumulative histogram of the pair separation $|\rvec_{12}|$ (nearest image) for 
two charges in a periodic three-dimensional cubic simulation box with periodic 
boundary 
conditions ($\beta c_1 c_2 = 2$, $L=1$).
\onecircle:
Reversible Markov-chain Monte Carlo (see~\cite[Fig. 8]{Faulkner2018})
\twocircle:
Method of
\sect{sec:CookbookAtomsNaive}
\threecircle:
Method of
\sect{sec:CookbookAtomsCell}
\fourcircle:
Method of
\sect{sec:CookbookAtomsVeto}
, each with 
standard errors for $\pi(|\rvec_{12}| < 0.6)$.}
\label{fig:Figure8abcConfirmation}
\end{figure}

The configuration file's \SECTION{InputOutputHandler} section specifies the 
input-output handler. It consists of the separation-output handler (an instance 
of the \CLASS{SeparationOutputHandler} class), which is connected to the 
\TAG{sampling} event handler. In the present example, it samples the 
nearest-image separation  (under periodic boundary conditions) of any two \pps. 
The initial global physical state is created randomly by the random-input 
handler (an instance of the \CLASS{RandomInputHandler} class). The configuration 
file \INI{coulomb_atoms/power_bounded.ini} reproduces published data (see 
\subfig{fig:Figure8abcConfirmation}{,\twocircle}).

The configuration file \INI{coulomb_atoms/power_bounded.ini} can be modified  
for $N$ \pps. In the \SECTION{RandomInputHandler} section, 
the number of root nodes must then equal $N$. 
In the \SECTION{Coulomb} 
section, the number of event handlers must be set to at least $N-1$ (this 
instructs the \TAG{Coulomb} tagger to deep-copy the required number of 
event handlers). Without changing the factor-type map with respect to the $N=2$ 
case, each event handler which will be presented with the correct in-state 
corresponding to a pair of units with one of them being the active unit.
The complexity of the implemented algorithm is \bigO{N} per event. 

\subsubsection{Atomic factors, cell-based bounding potential}
\label{sec:CookbookAtomsCell}

The configuration file \INI{coulomb_atoms/cell_bounded.ini} implements a single 
Coulomb pair factor with the merged-image Coulomb potential, just as the 
configuration file of \sect{sec:CookbookAtomsNaive}. However, a cell-occupancy
internal state associates the 
factor potential with a cell-based bounding potential. The target (non-active) 
unit may be 
cell-based or surplus (see \fig{fig:CellSystem3D}). The target unit may also be 
in an excluded nearby cell of the active cell (see \fig{fig:CellMethods}), for 
which the cell-based 
bounding potential cannot be used. In consequence, three taggers correspond to 
distinct event handlers that together realize the Coulomb pair factor. The 
consistency requirement of \JFVS assures that particles and units are always 
associated with the same cell.

Taggers and their tags are listed in the \SECTION{TagActivator} section. The 
\TAG{coulomb_cell_bounding} tagger, for example,  appears as an instance of the 
\CLASS{CellBoundingPotentialTagger} class. 
The \TAG{coulomb_cell_bounding} event 
handler then realizes the Coulomb factor unless the cell of the target particle 
 is excluded with respect to  the active cell and unless it is a surplus 
particle (in these cases the tagger does not generate any in-state for its 
event handler). Otherwise, the Coulomb pair factor is realized by a 
\TAG{coulomb_surplus}-tagged event handler or by a \TAG{coulomb_nearby} event 
handler. (For two units, as the active unit is taken out of the 
cell-occupancy system, no surplus candidate events are ever created.) 

The cell-occupancy systems (an instance of the \CLASS{SingleActiveCellOccupancy} 
class) is also declared in the \SECTION{TagActivator} section and further 
specified in the \SECTION{SingleActiveCellOccupancy}  section. The associated 
cell system  is described in the  \SECTION{CuboidPeriodicCells} section. The 
internal state, set in the \SECTION{SingleActiveCellOccupancy} section, has no 
charge value. This indicates that the identifiers of all particles at the cell 
level (here \pythoninline{cell_level} $= 1$) are tracked (see 
\sect{sec:CookbookSPCMolecularCB} for an example where this is handled 
differently).

The \TAG{coulomb_nearby} tagger, an instance of the \CLASS{ExcludedCellsTagger} 
class, yields the identifiers of particles in excluded cells of the active cell, 
by iterating over 
cells and by checking whether they contain appropriate identifiers. The 
\TAG{coulomb_surplus} tagger similarly relies on the \METHOD{yield_surplus} 
method of the cell-occupancy system to generate in-states.

To keep the internal state consistent with the global state, a cell-boundary 
event handler is used in the \CLASS{CellBoundaryTagger} class (together, this 
builds \TAG{cell_boundary} candidate events). The cell-boundary tagger just 
yields the active-unit identifier as the in-state 
used in the corresponding event handler. 
The configuration file \INI{coulomb_atoms/cell_bounded.ini} reproduces 
published data (see \subfig{fig:Figure8abcConfirmation}{,\threecircle}).

To adapt the configuration file for $N > 2$ \pps (from the $N=2$ case that is 
provided), in the 
\SECTION{RandomInputHandler}, \PROPERTY{number_of_root_nodes} must be set to 
$N$. The number of \TAG{coulomb_cell_bounding}, \TAG{coulomb_nearby}, and  
\TAG{coulomb_surplus} event handlers must be increased. Surplus particles can 
now exist. The number of event handlers to allow for depends on the 
cell 
system, whose parameters must be adapted in order to limit the number of 
surplus particles, and also to retain useful cell-based bounds 
for the Coulomb event rates.

\subsubsection{Atomic factors, cell-veto}
\label{sec:CookbookAtomsVeto}

The configuration file \INI{coulomb_atoms/cell_veto.ini} implements a 
Coulomb pair factor together with the merged-image Coulomb potential.
A cell-occupancy internal state is used. The
Coulomb pair factor is then realized, among others, by a cell-veto event 
handler, which associates the merged-image Coulomb potential with a cell-based 
bounding potential.

All the Coulomb pair factors of the active particle with target particles that 
are neither excluded nor surplus are taken together 
in a set of Coulomb factors, and realized by a single \TAG{coulomb_cell_veto}  
event handler. The candidate event-time can be calculated with the branch of the 
active unit 
as the in-state, which is implemented in the \CLASS{CellVetoTagger} class. 
(The 
cell-veto tagger returns the identifier of the active unit.)
The event handler returns the target 
cell (in which the candidate unit is to be localized)
together with the candidate event time.
The out-state request is accompanied by the branch of the target unit (if it 
exists), and the out-state computation is in analogy with the case 
studied in \sect{sec:CookbookAtomsCell}.

The configuration file features the \TAG{coulomb_cell_veto} tag together with 
the \TAG{coulomb_nearby}, \TAG{coulomb_surplus}, \TAG{cell_boundary}, 
\TAG{sampling}, \TAG{end_of_chain}, \TAG{start_of_run}, and \TAG{end_of_run} 
tags. The configuration file reproduces 
published data (see \subfig{fig:Figure8abcConfirmation}{,\fourcircle}).

To adapt the configuration file for $N$ \pps, the number of root nodes must be 
set to $N$ in the 
\SECTION{RandomInputHandler} section. The 
number of event handlers for the \TAG{coulomb_nearby} and  
\TAG{coulomb_surplus} 
events might have to be increased. However, a single cell-veto event handler 
realizes any number of factors with cell-based target particles whereas in 
\sect{sec:CookbookAtomsCell} each of them required its own event handler.

\subsection{Interacting dipoles}
\label{sec:CookbookDipoles}

\begin{figure}[htb]
\begin{center}
\includegraphics{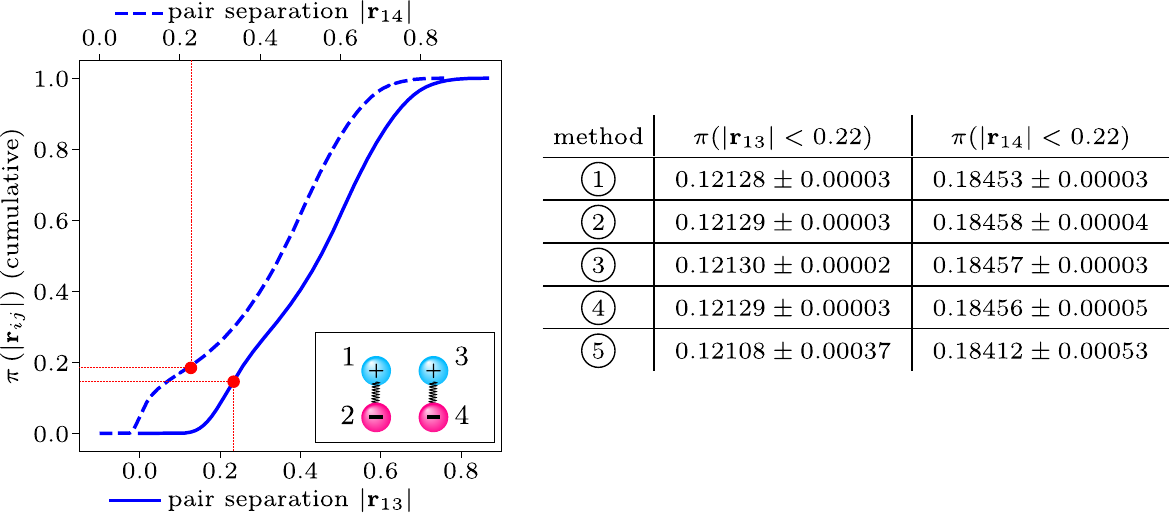}
\end{center}
\caption{
Cumulative histogram of the pair separation $|\rvec_{13}|$ and $|\rvec_{14}|$
(nearest image) for 
two dipoles (see the inset) in a periodic three-dimensional 
cubic simulation box with periodic boundary 
conditions ($\beta c_i c_j = \pm 1$, $L=1$).
\onecircle:
Reversible Markov-chain Monte Carlo (see~\cite[Fig. 11]{Faulkner2018})
\twocircle:
Method of
\sect{sec:CookbookDipolesLeaves}
\threecircle:
Method of
\sect{sec:CookbookDipolesCellBounded}
\fourcircle:
Method of
\sect{sec:CookbookDipolesCellVeto}
\fivecircle:
Method of
\sect{sec:CookbookDipolesLeavesRoots}, each with 
standard errors for $\pi(|\rvec_{13}| < 0.22)$ and $\pi(|\rvec_{14}| < 0.22)$.}
\label{fig:Figure11Validation}
\end{figure}

The configuration files in the \pythoninline{dipoles} directory of \JFVS 
implement the ECMC sampling of the Boltzmann distribution for two identical 
finite-size dipoles, for a model that  was introduced 
previously~\cite{Faulkner2018}. \PPS in different dipoles interact \via the 
merged-image Coulomb potential (pairs $1-3$, $1-4$, $2-3$, $2-4$ in 
\fig{fig:Figure11Validation}). \PPS within each dipole interact with a 
short-ranged potential (pairs $1-2$ and $3-4$). A repulsive short-range 
potential between oppositely charged atoms in different dipoles counterbalances 
the attractive Coulomb potential at small distances (pairs $1-4$ and $2-3$). 

Each dipole is a \cpp made up of two oppositely charged \pps. It is represented 
as a tree with one root node that has two children. The number of root nodes in 
the system is set in the \SECTION{RandomInputHandler} section of the 
configuration file, where the dipoles are created randomly through the 
\METHOD{fill_root_node} method in the \CLASS{DipoleRandomNodeCreator} class.
In the \PACKAGE{setting} package, 
the input handler specifies that there are two root 
nodes (\pythoninline{number_of_root_nodes} $=2$). Each of them contains two 
nodes (which is coded as \pythoninline{number_of_nodes_per_root_node}  $= 2$) 
and the number of 
node levels is two (\pythoninline{number_of_node_levels} $= 2$). As this 
numbers are set in the \PACKAGE{setting} package, 
all the \JF modules can autonomously construct all possible particle 
identifiers.

Statistically equivalent output is obtained for pair 
factors for all interactions (\sect{sec:CookbookDipolesLeaves}), for 
dipole--dipole Coulomb factors and their factor potential associated with a 
cell-based bounding potential 
(\sect{sec:CookbookDipolesCellBounded}), for dipole--dipole Coulomb factors 
with 
the cell-veto algorithm (\sect{sec:CookbookDipolesCellVeto}), and by
alternating between concurrent moves of 
the entire dipoles with moves of the individual \pps 
(\sect{sec:CookbookDipolesLeavesRoots}). The latter example showcases 
the collective-motion possibilities of ECMC integrated into \JF. 
All configuration files here implement the short-ranged potential 
as an instance of the \CLASS{DisplacedEvenPowerPotential} class with 
\pythoninline{power} $=2$ and the repulsive short-range potential as an 
instance of the \CLASS{InversePowerPotential} class with \pythoninline{power} 
$=6$.

\subsubsection{Atomic Coulomb factors}
\label{sec:CookbookDipolesLeaves}
The configuration file \INI{dipoles/atom_factors.ini} implements for each 
concerned pair of \pps a Coulomb pair factor, with the merged-image Coulomb 
potential associated with the the inverse-power Coulomb bounding potential. 
Several event handlers that are 
instances of the same class realize these factors, and the number of event 
handlers must scale with their number. No internal state is declared. Pair 
factors are implemented for each pair of \pps that interact with a harmonic or 
a 
repulsive potential. One of the four \pps is active at each time, and it moves 
either in $+x$, in $+y$, or in $+z$ direction. 
The configuration file represents \cpps as trees with two levels 
(see \sect{sec:IntroductionGlobalInternal}). 
Positions and velocities are kept consistent on both levels, although the 
root-unit properties are not made use of. The tree structure only serves to 
identify leaf units on the same dipole.

In the configuration file, taggers and tags are listed in the 
\SECTION{TagActivator} section.
The \TAG{coulomb}, \TAG{harmonic}, and  
\TAG{repulsive} taggers are separate instances of the same
\CLASS{FactorTypeMapInStateTagger} class, and the corresponding 
sections set up the corresponding event handlers. Both the \TAG{harmonic} and 
the \TAG{repulsive} event handlers are instances of the 
\CLASS{TwoLeafUnitEventHandler} class. Aliasing nevertheless assures a
tree-structured configuration file (the \TAG{harmonic} tagger 
is for example declared with a \CLASS{HarmonicEventHandler} class 
which is an alias for the \CLASS{TwoLeafUnitEventHandler} class).
The \TAG{coulomb} tagger and its event handlers are treated  as in 
\sect{sec:CookbookAtomsNaive}.

The sampling, start-of-run, end-of-run and end-of-chain pseudo-factors are 
realized by event handlers that are set up in the same way as in all other 
configuration files. However, the parent sections differ: the parent of the 
\SECTION{InitialChainStartOfRunEventHandler} section sets the 
\TAG{start_of_run} tagger, which specifies that after the \TAG{start_of_run} 
event, new \TAG{coulomb}, \TAG{harmonic}, \TAG{repulsive}, \TAG{sampling} 
\TAG{end_of_chain}, and \TAG{end_of_run} event handlers must be activated. The 
tag lists thus differ from those of the \SECTION{StartOfRun} section in other 
configuration files. The configuration file \INI{dipoles/atom_factors.ini} 
reproduces published data (see \subfig{fig:Figure11Validation}{,\twocircle}).

\subsubsection{Molecular Coulomb factors, cell-based bounding 
potential } 
\label{sec:CookbookDipolesCellBounded}

The configuration file \INI{dipoles/cell_bounded.ini} implements for each pair 
of dipoles a Coulomb four-body factor. (The sum of the merged-image Coulomb 
potentials for pairs $1-3$, $1-4$, $2-3$, $2-4$  in \fig{fig:Figure11Validation} 
constitutes the Coulomb factor potential.) The event rates for such factors 
decay much faster with distance than for Coulomb pair factors, and the 
chosen lifting scheme considerably influences the dynamics (see~\cite[Sect. 
IV]{Faulkner2018}). The configuration file installs a cell-occupancy  internal 
state on the dipole level (rather than for the \pps). A cell-bounded event 
handler then realizes a Coulomb four-body factor with its factor potential 
associated with an orientation-independent cell-based bounding potential for 
dipole pairs that are not in excluded cells relative to each other.  The 
configuration file furthermore implements pair factors for the  harmonic and the 
 repulsive interactions. One of the four \pps is active at each time, and it 
moves either in $+x$, in $+y$, or in $+z$ direction. 

The configuration file's \SECTION{TagActivator} section defines all taggers and 
their corresponding tags. Among the taggers for event handlers realizing the 
Coulomb four-body factor, the \TAG{coulomb_cell_bounding} tagger differs 
markedly from the set-up in \sect{sec:CookbookAtomsCell}, as the event 
handler\footnote{set in the 
\SECTION{TwoCompositeObjectCellBoundingPotentialEventHandler} section} is for a 
pair of \cpps. The lifting scheme is set to \pythoninline{inside_first_lifting}. 
The bounding potential is defined in the \SECTION{CellBoundingPotential} 
section. A dipole Monte Carlo estimator is used for simplicity (see 
\sect{sec:SystemEstimator}). As it obtains an upper bound for the event rate 
from  random trials for each relative cell orientations, its use is restricted 
to there being only a small number of cells. The \TAG{coulomb_nearby} and 
\TAG{coulomb_surplus} taggers are for event handlers realizing the Coulomb 
four-body factor when the bounding potential cannot be used. In this case, the 
merged-image Coulomb potential is summed for the factor potential, but also for 
the bounding potential.\footnote{The tree structure of the configuration file is 
hidden in this case, as the \JF factory (which builds instances of classes 
based 
on its content) creates separate instances for all the descendants of a 
section, not requiring the use of aliases.} The standard \TAG{sampling}, 
\TAG{end_of_chain}, \TAG{end_of_run}, \TAG{start_of_run} taggers as well as the 
ones responsible for the harmonic and repulsive potentials are set up in a 
similar way as in \sect{sec:CookbookDipolesLeaves}.

The \SECTION{TagActivator} section defines the internal state that is used by 
the \TAG{coulomb_cell_bounding}, \TAG{coulomb_nearby}, and 
\TAG{coulomb_surplus} 
taggers. The \SECTION{SingleActiveCellOccupancy} 
section specifies the cell level (\PROPERTY{cell_level}  $= 1$ indicates 
that the particle identifiers have length one, corresponding to root nodes, 
rather than length two, which would correspond to the dipoles' leaf nodes). 
Positions and velocities must thus be kept consistent on both levels.
The 
cell-occupancy system requires the presence of a \TAG{cell_boundary} event 
handler,  again 
on the level of the root nodes. This event handler is aware of the cell level, 
and it ensures consistency of the events triggered by the cell-based bounding 
potential with the underlying cell system.
The configuration file \INI{dipoles/cell_bounded.ini} reproduces 
published data (see \subfig{fig:Figure11Validation}{,\threecircle}).

\subsubsection{Molecular Coulomb factors, cell-veto} 
\label{sec:CookbookDipolesCellVeto}

The configuration file \INI{dipoles/cell_veto.ini} implements the same factors 
and pseudo-factors and the same internal state as the configuration file of 
\sect{sec:CookbookDipolesCellBounded}. A single cell-veto event handler then 
realizes the set of factors that relate to cells that are not excluded for any 
number of cell-based particles, whereas in the earlier implementation, the 
number of  cell-bounded event handlers must exceed the possible number of 
particles in non-excluded cells of the active cell. This is what allows to 
implement ECMC with a 
complexity of \bigO{1} per event. 

The configuration file resembles that of \sect{sec:CookbookDipolesCellBounded}. 
It mainly replaces the latter file's \TAG{coulomb_cell_bounding} event handlers 
with a \TAG{coulomb_cell_veto} event handler. Slight differences reflect 
the 
fact that a cell-veto event handler uses no \METHOD{displacement} method 
of the bounding potential but obtains the displacement from the total event 
rate (see the discussion in \sect{sec:EventHandlersPairCellVeto}).
The configuration file \INI{dipoles/cell_veto.ini} reproduces 
published data (see \subfig{fig:Figure11Validation}{,\fourcircle}).

\subsubsection{Atomic Coulomb factors, alternating root mode and leaf mode}
\label{sec:CookbookDipolesLeavesRoots}

\begin{figure}[htb]
\begin{center}
\includegraphics{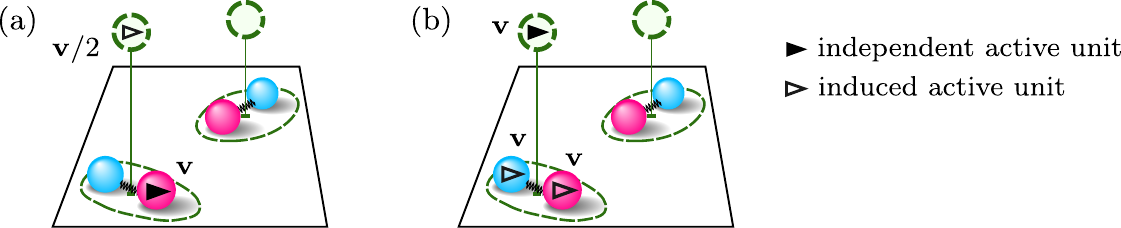}
\end{center}
\caption{Two moves implemented in 
\protect \INI{dipoles/dipole_motion.ini}.
\subcap{a} In leaf mode, a single independent active leaf unit  has velocity 
$\vvec$. The corresponding dipole center (the active root unit) is induced to 
move at $\vvec/2$. 
\subcap{b} In root mode, one dipole (independent active root unit) has
velocity 
$\vvec$, and both its active leaf units have induced velocity $\vvec$.}
\label{fig:TwoDipoleModes}
\end{figure}

The configuration file \INI{dipoles/dipole_motion.ini} implements two different 
modes. In leaf mode, at each time one of the four \pps is active, 
and it moves 
either in $+x$, in $+y$, or in $+z$ direction (see 
\subfig{fig:TwoDipoleModes}{a}). In root mode, at each 
time the \pps of one
dipole moves as a rigid block, in the same direction (see 
\subfig{fig:TwoDipoleModes}{b}). (The root mode, by itself, does not assure 
irreducibility of the Markov-chain algorithm, as the orientation and shape of 
any dipole molecule would remain unchanged throughout the run.)

\JFVS represents the dipoles as trees, and both 
modes are easily implemented. In leaf mode, the Coulomb factors are 
realized by \TAG{coulomb_leaf} event handlers that are instances of the same 
class\footnote{Instances of the 
\CLASS{TwoCompositeObjectSummedBoundingPotentialEventHandler} class} as the   
\TAG{coulomb_nearby} event handlers 
in \secttwo{sec:CookbookDipolesCellBounded}{sec:CookbookDipolesCellVeto}.
The root mode, in turn, is patterned 
after the simulation of two \pps (as in 
\sect{sec:CookbookAtomsNaive}): all inner-dipole potentials are constant. The 
inter-dipole Coulomb 
potentials sum up to an effective two-body potential, the factor potential of a 
two-body factor realized in a Coulomb-dipole event handler.
The repulsive 
short-range potential between oppositely charged atoms in different dipoles 
also 
translates into a potential between the dipoles in rigid motion, and serves as 
a factor potential of a two-body factor, realized in a specific event handler. 

Taggers and their tags are listed in the \SECTION{TagActivator} section.
The \TAG{harmonic_leaf}, \TAG{repulsive_leaf} (leaf-mode) taggers, as 
well as all those related to event handlers that realize pseudo-factors are as 
in \sect{sec:CookbookDipolesLeaves}. The \TAG{coulomb_leaf} tagger corresponds 
to the 
\TAG{coulomb_nearby} 
tagger in \sect{sec:CookbookDipolesCellBounded}. 
The \TAG{coulomb_root} and \TAG{repulsive_root} taggers are analogous to those 
in \sect{sec:CookbookAtomsNaive} for the two-atom case.

As all other operations that take place in \JF, the switches between leaf mode 
and root mode are also formulated as events. They are related to two 
pseudo-factors and 
realized 
by a \TAG{leaf_to_root} event handler
and by a \TAG{root_to_leaf} event handler, respectively.
(These two event handlers are aliases for instances of the
\CLASS{RootLeafUnitActiveSwitcher} class.)
The \TAG{root_to_leaf} and 
\TAG{leaf_to_root} taggers, in addition to the \pythoninline{create} and 
\pythoninline{trash} lists, set up separate \pythoninline{activate} and 
\pythoninline{deactivate} lists (see \sect{sec:ArchitectureActivator}).
The configuration file reproduces 
published data (see \subfig{fig:Figure11Validation}{,\fivecircle}).
Of particular interest is that the tree representation of \cpps keeps 
consistency between leaf-node units and root-node units: the event handlers 
return branches of cnodes for all independent units (see 
\fig{fig:StateHandler}) whose unit information can be 
integrated into the global state. 

\subsection{Interacting water molecules (SPC/Fw model)}
\label{sec:CookbookSPC} 

The configuration files in the \pythoninline{water} directory 
implement the ECMC sampling of the Boltzmann distribution for 
two water molecules, using
the SPC/Fw model that was previously studied with
ECMC~\cite{Faulkner2018}. Molecules are represented as \cpps 
with three charged \pps, one of which is positively charged (representing the 
oxygen) and the two others are negatively charged (representing the 
hydrogens). \PPS in different water molecules interact \via the merged-image 
Coulomb potential. In addition, \pps within each molecule interact with a 
three-body bending interaction, and a harmonic oxygen--hydrogen potential. 
Finally, any two oxygens interact through a Lennard-Jones 
potential~\cite{Faulkner2018}. 

In the tree state handler (defined in the \SECTION{TreeStateHandler} section, 
a child of the \SECTION{SingleProcessMediator} section), 
water molecules are represented as trees with a root node and 
three children (the leaf nodes of the tree). The total number of water 
molecules (that is, of root nodes) is set in the \SECTION{RandomInputHandler} 
section of each configuration file. The molecules are created through the 
\METHOD{fill_root_node} method in  the \CLASS{WaterRandomNodeCreator} class. 
There are two node levels (\pythoninline{number_of_node_levels} $= 2$) 
and three nodes per root node (\pythoninline{number_of_nodes_per_root_node} 
$=3$). The charges of a molecule are set in the 
\SECTION{ElectricChargeValues} section 
(a descendant of the \SECTION{WaterRandomNodeCreator} section).

All the configuration files in the \pythoninline{water} directory of \JFVS 
implement the pair harmonic factors that are realized through \TAG{harmonic} 
event handlers. The corresponding taggers are defined in the \SECTION{Harmonic} 
sections, with the displaced even-power potential and its parameters set
in the \SECTION{HarmonicEventHandler} and \SECTION{HarmonicPotential} sections. 
The configuration files furthermore implement the taggers corresponding to the 
three-body bending factors in their \SECTION{Bending} sections. The 
\TAG{bending} event handler has three independent units (attached to branches). 
It thus requires a lifting scheme (which is chosen in the 
\SECTION{BendingEventHandler} 
section), which is however unique (see~\cite[Fig. 2]{Faulkner2018}). In all 
these configuration files, one of the six \pps is active, and it moves either in 
$+x$, in $+y$, or in $+z$ direction (the optional rigid displacement of the 
entire water molecule, could be set up as in 
\sect{sec:CookbookDipolesLeavesRoots}).

Statistically equivalent output is obtained for a simple set-up featuring pair 
factors for the Coulomb  potential and a Lennard-Jones interaction that is 
inverted (\sect{sec:CookbookSPCAtomicInv}), or  for a molecular-factor Coulomb 
potential associated with a power-law bounding potential and a cell-based 
Lennard-Jones bounding potential (\sect{sec:CookbookSPCMolecularCB}). In 
addition, the cell-veto algorithm for the Coulomb potential coupled to an 
inverted Lennard-Jones potential (\sect{sec:CookbookSPCCellInv}) is also 
provided. Finally, cell-veto event handlers take part in the realization of 
complex molecular Coulomb factors  and also realize Lennard-Jones factors 
between oxygens (\sect{sec:CookbookSPCCVCV}). This illustrates how multiple 
independent cell-occupancy systems may coexist within the same run.

\begin{figure}[htb]
\begin{center}
\includegraphics{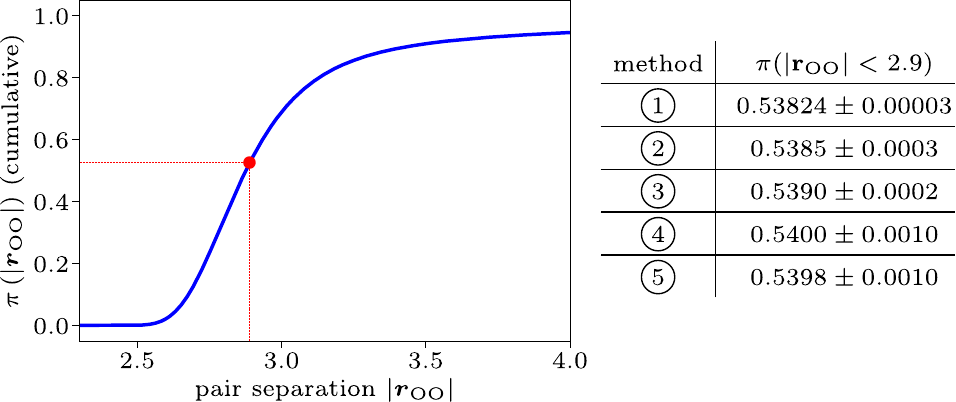}
\end{center}
\caption{Cumulative histogram of the oxygen--oxygen pair separation 
$|\rvec_{\text{OO}}|$
for two SPC/Fw water molecules 
in a periodic
cubic simulation box. 
\onecircle:
Reversible Markov-chain Monte Carlo
(see~\cite[Fig. 14]{Faulkner2018})
\twocircle:
Method of
\sect{sec:CookbookSPCAtomicInv}
\threecircle:
Method of
\sect{sec:CookbookSPCMolecularCB}
\fourcircle:
Method of
\sect{sec:CookbookSPCCellInv}
\fivecircle:
Method of \sect{sec:CookbookSPCCVCV}, each with 
standard errors for $\pi(|\rvec_{\text{OO}}| < 2.9)$.}
\label{fig:Figure14Validation}
\end{figure}

\subsubsection{Atomic Coulomb factors, Lennard-Jones inverted}
\label{sec:CookbookSPCAtomicInv}

The configuration file \INI{water/coulomb_power_bounded_lj_inverted.ini} 
implements pair Lennard-Jones, harmonic and Coulomb factors. The Coulomb 
factors are realized for any distance of the point masses by event handlers 
that associate the merged-image Coulomb potential with its inverse-power 
Coulomb 
bounding potential. 
The Lennard-Jones potential is inverted.
This configuration file needs no internal state.

In the configuration file, the \SECTION{TagActivator} section lists all the 
taggers together with their tags, which in addition to the taggers related to 
pseudo-factors, are reduced to \TAG{coulomb}, \TAG{harmonic}, 
\TAG{bending}, and \TAG{lennard_jones}.
The merged-image Coulomb potential with its associated power-law bounding 
potential (both for attractive and repulsive charge products) is specified in 
the \SECTION{Coulomb} section of the configuration file.
The Lennard-Jones potential is invertible and its 
\METHOD{displacement} method is used rather than that of a bounding potential.
The output handler is defined in the 
\SECTION{OxygenOxygenSeparationOutputHandler} section, a child of the 
\SECTION{InputOutputHandler} section. It obtains all the units, extracts the 
oxygens through their unit identifier, and records the oxygen--oxygen 
separation distance. 
This reproduces published data (see 
\subfig{fig:Figure14Validation}{,\twocircle}).

\subsubsection{Molecular Coulomb factors, Lennard-Jones cell-bounded}
\label{sec:CookbookSPCMolecularCB}

The configuration file \INI{water/coulomb_power_bounded_lj_cell_bounded.ini} for 
the water system corresponds to pair factors for the Lennard-Jones and the 
harmonic potentials and to molecular factors for the Coulomb interaction. The 
Coulomb factor potential is the sum of the merged-image Coulomb potential for 
the nine relevant pairs of \pps (pairs across two molecules). It is realized in 
a particular event handler,\footnote{an instance of the 
\CLASS{TwoCompositeObjectSummedBoundingPotentialEventHandler} class.} 
analogously to how this is done for the Coulomb interaction in 
\secttwo{sec:CookbookDipolesCellBounded}{sec:CookbookDipolesCellVeto}. The 
associated bounding potential (both for attractive and repulsive charge 
combinations) is given by the sum over all the individual pairs. Although the 
Lennard-Jones interaction can be inverted, the configuration file sets up a 
cell-occupancy internal state that tracks the identifiers for the oxygens. As in 
previous cases, this leads to three types of events, corresponding to the 
nearby, surplus, and cell-based particles, in addition to cell-boundary events.

Taggers and their tags are listed in the \SECTION{TagActivator} section. Taggers 
are generally utilized as in other configuration files. The internal state is 
specified in the \SECTION{TagActivator} section. As set up in the 
\SECTION{SingleActiveCellOccupancy} section, it features a 
\pythoninline{oxygen_indicator} charge (set in the \SECTION{OxygenIndicator} 
section). The oxygen-indicator charge is non-zero only for the oxygens.    In 
consequence, the oxygen cell system (defined in the \SECTION{OxygenCell} 
section) tracks only  oxygens. This reproduces published data (see 
\subfig{fig:Figure14Validation}{,\threecircle}). Nevertheless, this 
configuration file does not scale up easily with system size. 

\subsubsection{Molecular Coulomb cell-veto, Lennard-Jones inverted } 
\label{sec:CookbookSPCCellInv}

The configuration file \INI{water/coulomb_cell_veto_lj_inverted.ini} for the 
water system corresponds to the same factors as in 
\sect{sec:CookbookSPCMolecularCB}. As a preliminary step towards the treatment 
of all long-range interactions with the cell-veto algorithm, in 
\sect{sec:CookbookSPCCVCV}, molecular Coulomb factors are realized here (for 
non-excluded cells of the active cell) with a cell-veto event handler.

Taggers and their tags are listed in the \SECTION{TagActivator} section, 
and they are generally similar to those of other configuration files. 
In addition,  the internal state for the Coulomb system  is defined in the 
\SECTION{TagActivator} 
section and
further described in the \SECTION{SingleActiveCellOccupancy}
section.
The latter 
describes the cell level (which serves for the water molecules) as on the 
root node level 
(\pythoninline{cell_level} $=1$), the barycenter  of the leaf-node positions
of each water molecule. (Root-node and leaf-node positions are set in the 
random input handler, which itself uses a water random node creator.) 

The event handlers consistently update all leaf-node positions and root-node 
positions from a valid initial configuration obtained in an instance of 
the \CLASS{WaterRandomNodeCreator} class. Consistency will be deteriorated over 
long runs, but this is of little importance for the simple example case 
presented here. 
The configuration file reproduces published data (see 
\subfig{fig:Figure14Validation}{,\fourcircle}).

\subsubsection{Molecular Coulomb cell-veto, Lennard-Jones cell-veto}
\label{sec:CookbookSPCCVCV}

The configuration file \INI{water/coulomb_cell_veto_lj_cell_veto.ini} offers no 
new factors compared to 
\secttwo{sec:CookbookSPCMolecularCB}{sec:CookbookSPCCellInv}, but it uses, for 
illustration purposes, two cell-occupancy systems and two cell-veto event 
handlers. As nearby and surplus particles are excluded from the cell-veto 
treatment, this implies two sets of cell-based, nearby, and surplus event 
handlers in addition to two cell-boundary event handlers. For the molecular 
Coulomb factors, the cell-veto event handler receives as a factor potential the 
sum of pairwise merged-image Coulomb potentials with attractive and repulsive 
charge combinations. Cells track the barycenter of individual water molecules, 
and consistency between root-node units and leaf-node units is of importance. 
Although the Lennard-Jones potential can be inverted, the configuration file 
sets up a second cell-occupancy system for the Lennard-Jones potential. The 
cell-occupancy system tracks only 
leaf-node particles that correspond to oxygen atoms. 

Taggers and their tags are listed in the 
\SECTION{TagActivator} section. This section is of interest as it sets up the 
internal state as two cell-occupancy systems, both instances of the same  
\CLASS{SingleActiveCellOccupancy} class. They require different parameters, and 
are therefore presented under aliases, in the \SECTION{OxygenCell} and 
\SECTION{MoleculeCell} sections. Each of theses cell-occupancy systems use a 
separate cell system instance of the same class. As the two cell systems have 
the same parameters, they do not need to be aliased in the configuration file. 
The configuration file reproduces published data (see 
\subfig{fig:Figure14Validation}{,\fivecircle}).

\section{Licence, GitHub repository, Python version}
\label{sec:LicGitHubPython}

\JF, the Python application described in this paper, is an
open-source  software
project that grants users the rights to study and execute, modify and 
distribute the code. Modifications can be fed back into the project. 

\subsection{Licence information, used software}
\label{sec:LicGitHubPythonLicence}

\JF is made available under the GNU GPLv3 licence (for details see the 
\JF\ \PATH{LICENCE} file). The use of the Python \PACKAGE{MDAnalysis} 
package~\cite{MichaudAgrawal2011,Gowers2016} for reading and writing 
\PATH{.pdb} files, 
of the Python \PACKAGE{Dill} package~\cite{Mckerns2010,Mckerns2011} for dumping 
and restarting a run of the application,
and of the Python \PACKAGE{Matplotlib}~\cite{Hunter2007} and 
\PACKAGE{NumPy}~\cite{Oliphant2006,Walt2011} 
packages for the 
graphical analysis of output is acknowledged.

\subsection{GitHub repository}
\label{sec:LicGitHubPythonGitHub}
\PATH{JeLLyFysh}, the public repository for all the 
code 
and the documentation of the application, is part of a public GitHub 
organization.\footnote{The organization's url is  
\url{https://github.com/jellyfysh}}
The repository can be forked (that is, 
copied to an outside user's own public repository) and from there studied, 
modified and run in the user's local environment. Users may
contribute to the \JFA  \via pull requests (see the 
\JF\ \PATH{README} and \PATH{CONTRIBUTING.md} files for 
instructions and guidelines). All communication (bug reports, suggestions) 
takes place through 
GitHub \quot{Issues}, that can be opened in the repository by any user or 
contributor, and that are classified in GitHub projects on 
\PATH{JeLLyFysh}.

\subsection{Python version, coding conventions}
\label{sec:LicGitHubPythonPython}
\JFVS is compatible with Python~3.5 (and higher) and with 
PyPy~7 (and higher), a just-in-time compiling Python  alternative to 
interpreted  CPython (see the \JF documentation for details). \JF code adheres 
to the PEP8 style guide for Python code, except for the linewidth that is set 
to 120 (see the \PATH{CONTRIBUTING.md} file for details).

Following the PEP8 Python naming convention, \JF modules and 
packages are spelled in snake case and classes in camel case (the 
\MODULE{state_handler} module thus contains the \CLASS{StateHandler} 
class). In configuration files, section titles are in camel case and enclosed 
in 
square brackets (see \fig{fig:ConfigFileTree}).

Versioning of the \JF project  adopts two-to-four-field version numbers defined 
as Milestone.Feature.AddOn.Patch. Version 1.0, as described, represents the 
first development milestone which reproduces published
data~\cite{Faulkner2018}.
Patches and bugfixes of this version will be given number 
1.0.0.1, 1.0.0.2, etc. (Finer-grained distinction between versions is obtained 
through the hashes of master-branch GitHub commits.) New configuration files 
and 
required extensions are expected to lead to versions 1.0.1. 1.0.2, etc. Version 
1.1 is expected to fully implement different dimensions 
and arbitrary rectangular and cuboid shapes of the \JF\ \PACKAGE{potential} 
package. Versions 1.2 and 1.3 will consistently implement $\NACT > 1$ 
independent active particles (on a single processor) and eliminate unnecessary
time-slicing for some events triggered by pseudo-factors and for unconfirmed 
events.
All development from Versions 1.0 to 2.0 
can be undertaken concurrently. Fully parallel code is planned for Version 3.0. 
In \JF development, two-field versions (2.0, 3.0, etc) may introduce 
incompatible code, while three- and four-field version numbers are intended to 
be backward compatible.

\section{Conclusions, outlook}
\label{sec:Conclusions}

As presented in this paper, \JF is a computer application for ECMC simulations 
that is hoped to become useful for researchers in different fields of 
computational science. The \JFVS constitutes its first development milestone: 
built on the mediator design pattern, it systematically formulates the entire 
ECMC time evolution in terms of events, from the start-of-run to the end-of-run, 
including sampling, restarts (that is, end-of-chain), and the factor events. A 
number of configuration files validate \JFVS  against published test cases for  
long-range interacting systems~\cite{Faulkner2018}.

For \JFVS, consistency has been the main concern, and code has 
not yet been optimized. Also, the handling of exceptions remains 
rudimentary, although this is not a problem for the cookbook 
examples of \sect{sec:Cookbook}.

All the methods are written in Python. Considerable speed-up can 
certainly be obtained by rewriting time-consuming parts of the application 
in compiled languages, in particular
of the \PACKAGE{potential} package.
One of the principal limitations of \JFVS is that pseudo-factor-related  
and unconfirmed events are time-sliced, 
leading to superfluous trashing and 
re-activation of candidate events. 
Optimized bounding potentials for many-particle factor 
potentials appear also as a priority.

The consistent implementation of an arbitrary number $\NACT$ of simultaneously 
active particles is straightforward, although it has also not been implemented 
fully in \JFVS. (As mentioned, this is planned for \JFVTWO). This will enable 
full parallel implementations on multiprocessor machines. Simplified parallel 
implementations for  one-dimensional systems and for hard-disk models in two 
dimensions are currently being prototyped. The parallel computation of candidate 
events (using the \CLASS{MultiProcessMediator} class implemented in \JFVS)
is at present rather slow. Bringing the full power of parallelization and of 
multi-process ECMC to real-world applications appears as its outstanding 
challenge for \JF.

\section*{Acknowledgements}

P.H. acknowledges support from the Bonn-Cologne Graduate School of Phy\-sics and 
Astronomy honors branch and from Institut Philippe Meyer. L.Q. and M.F.F. 
acknowledge hospitality at the Max-Planck-Institut für Physik komplexer Systeme, 
Dresden, Germany. M.F.F. acknowledges financial support from EPSRC fellowship 
EP/P033830/1 and hospitality at Ecole normale sup\'{e}rieure.  W.K. acknowledges 
support from the Alexander von Humboldt Foundation.  

\bibliographystyle{eplbib.bst}
\bibliographystyle{elsarticle-harv}

\bibliography{General,PROG_2018}
\end{document}